\documentclass[a4paper,fleqn,usenatbib]{mnras}

\newcommand\bfv{\textbf{V}}

\usepackage[T1]{fontenc}
\usepackage{ae,aecompl}
\usepackage{bm}

\usepackage{graphicx}	
\usepackage{amsmath}	
\usepackage{amssymb}	
\usepackage{amsfonts}
\usepackage{verbatim}
\usepackage{mathrsfs}	
\usepackage{setspace}
\usepackage{mathrsfs}
\usepackage{bm}
\usepackage{verbatim}
\usepackage{xcolor}
\usepackage[multiple]{footmisc}

\usepackage{lipsum}

\title[Hydrodynamical instability in accretion discs]{Hydrodynamical instability with noise in the Keplerian accretion 
discs: Modified Landau equation}

\author[Ghosh \& Mukhopadhyay]{
Subham Ghosh,$^{1}$\thanks{subham@iisc.ac.in}
and Banibrata Mukhopadhyay$^{1}$\thanks{bm@iisc.ac.in}
\\
$^{1}$Department of Physics, Indian Institute of Science, Bangalore, Karnataka, India, 560012\\
}

\date{Accepted 2020 June 17. Received 2020 June 11; in original form 2020 April 7}

\pubyear{2020}

\begin{document}
\label{firstpage}
\pagerange{\pageref{firstpage}--\pageref{lastpage}}
\maketitle

\begin{abstract}
Origin of hydrodynamical instability and turbulence in the Keplerian accretion disc as well as similar laboratory 
shear flows, e.g. plane Couette flow, is a long standing puzzle. These flows are linearly stable. Here we explore the 
evolution of perturbation in such flows in the presence of an additional force. Such a force, which is expected to be 
stochastic in nature hence behaving as noise, could be result of thermal fluctuations (however small be), Brownian 
ratchet, grain-fluid interactions  and feedback from outflows in astrophysical discs etc. We essentially establish the 
evolution of nonlinear perturbation in the presence of Coriolis and external forces, which is modified Landau equation. 
We show that even in the linear regime, under suitable forcing and Reynolds number, the otherwise least stable 
perturbation evolves to a very large saturated amplitude, leading to nonlinearity and plausible turbulence. Hence, 
forcing essentially leads a linear stable mode to unstable. We further show that nonlinear perturbation diverges at a 
shorter timescale in the presence of force, leading to a fast transition to turbulence. Interestingly, emergence of 
nonlinearity depends only on the force but not on the initial amplitude of perturbation, unlike original Landau equation 
based solution.
\end{abstract}




\section{Introduction}

\label{sec:intro}

Accretion discs are ubiquitous in astrophysics in different forms. Examples are discs formed during birth of planetary 
systems, discs formed by the mass transfer from a companion object to the central denser object in binary systems, discs 
around the supermassive black holes at the center of galaxies. However, the process of transfer of matter inward and 
angular momentum outward is still not well understood due to the inadequate molecular viscosity of matter therein. Hence, 
to explain the observed luminosity (or temperature) from the disc, we must require other source of viscosity. It is 
generally believed that the turbulent viscosity helps in transporting angular momentum. \cite{1973A&A....24..337S} and 
later \cite{1974MNRAS.168..603L} prescribed the origin of turbulent viscosity in accretion discs, but rather in an ad hoc 
manner. The origin of turbulence was not uncovered then. There are pure hydrodynamical proposals to explain the angular 
momentum transport in accretion discs, mostly based on stability analysis and further turbulence. Some of these are: 
transient growth leading to nonlinearity in shear flows (\citealt{Lominadze_1988,Chageli_2003,
Tevzadze_2003,2005ApJ...629..383M,2005ApJ...629..373A,Shen_2006_apj,Lithwick_2007,Lithwick_2009}), the 
emergence of Rayleigh-Taylor type instability in the Keplerian flow due to the presence of vertical shear 
(\citealt{2013MNRAS.435.2610N, 2016A&A...586A..33U, 2015ApJ...811...17L, 2014A&A...572A..77S, 2016A&A...594A..57S, 
2015MNRAS.450...21B}), Zombie Vortex instability (\citealt{2013PhRvL.111h4501M, 2015ApJ...808...87M}), convective 
overstability (\citealt{2014ApJ...788...21K}), etc.  However they are not free from caveats. Also often they are 
insufficient to explain transport of angular momentum as inferred from observation, i.e. Shakura-Sunyaev viscosity 
parameter, $\alpha$ (\citealt{1973A&A....24..337S}), is quite small to explain the observations. Convective overstability 
has some saturation, it does not let the perturbation modes to grow indefinitely (\citealt{2016MNRAS.455.2608L}).

In the Magnetohydrodynamic (MHD) regime, \cite{1991ApJ...376..214B} found that the turbulence could be through the 
instability due to the interplay between magnetic field and rotation of the flow, following the idea of \cite{article} 
and \cite{1960PNAS...46..253C}. This instability is known as Magneto-Rotational Instability (MRI) and those authors 
showed that this linear instability in the presence of only weak magnetic field could give rise to  
MHD turbulence. MRI is extremely successful to explain the origin of turbulence in accretion discs over the years. 
However, it has some limitation too, particularly in the low ionization regime.
Although \cite{Salmeron_2004,Salmeron_2005,Salmeron_2008} argued for
the possible existence of MRI in colder accretion flows, particularly 
in the case of protoplanetary disc, based on ambipolar 
diffusion, Ohmic diffusion and Hall diffusion, 
they could not resolve the underlying dead zone problem in the accretion disc
completely. Indeed
\cite{Bai_2013_ApJ, Bai_2017, Bai_2013} showed through 
numerical simulations that due to the nonideal MHD effects, like ambipolar 
diffusion, Ohmic diffusion and Hall diffusion, MRI gets strongly affected,
which pose problem to explain protoplanetary discs.
The problem is particularly severe in the low states of cataclysmic variables 
(\citealt{Gammie_1998}), the outer part of disc in active galactic nuclei (AGNs) and 
the underlying dead zone (e.g. \citealt{menou,menq}), where the ionization 
is very small such that matter cannot be coupled with the 
magnetic field, hence MRI gets suppressed. It is, therefore, a general 
concern of the origin of hydrodynamic turbulence or instability 
leading to turbulence in these discs.

The limitations of MRI do not end here. \cite{2015PhRvE..92b3005N} showed that MRI may be suppressed beyond the Reynolds 
number ($Re$) $10^9$, unless perturbation is tuned appropriately, and at that regime it is the magnetic transient growth 
which brings nonlinearity and hence plausible turbulence in the system. Note that $Re$ in accretion discs is well above 
this critical value (\citealt{2013PhLB..721..151M}). 
Further, MRI is suppressed in the high resistive limit, while it is 
relevant only with specifically tuned perturbations in the ideal MHD limit.
Also in the ideal inviscid limit (i.e., $Re\rightarrow\infty$), apart from the exponential MRI growth at 
large times, the flow also undergoes transient growth during finite/dynamical times with comparable 
or higher growth factors, as demonstrated by \cite{Mamatsa_2013} (see also, \citealt{Bhatia_2016}).
Apart from this, \cite{2005ApJ...628..879P} showed that in compressible and differentially rotating 
flows, axisymmetric MRI gets stabilized beyond a toroidal component of the magnetic field. While their calculations were 
done in local approximation, \cite{2018MNRAS.473.2791D} confirmed the suppression of MRI in global analysis.

Nevertheless, there is a history of controversy about the stability of Rayleigh stable flows and hence the angular 
momentum transport via turbulent viscosity in these kind of flows, particularly in accretion discs, in the literature 
(e.g. \citealt{Dubrulle_2005_a, Dubrulle_2005_b, Dauchot_1995, Rudiger_2001, Klahr_2003, Richard_1999, Kim_2000, 
Mahajan_2008, Yecko_2004, Mukhopadhyay_Mathew_2011, Mukhopadhyay_Chattopadhyay_2013}). Efforts have been put forward to 
resolve this issue in the context of hot accretion discs by considering shearing sheet approximation, with (e.g. 
\citealt{Lesur_2005}) and without (e.g. \citealt{Balbus_1996, Hawley_1999}) viscosity. 
\cite{Fromang_2007}, based on MHD simulation, argued for the importance of dissipation, both
resistive and viscous, in order to conclude angular momentum transport and \cite{Pumir_1996} 
examined sustained turbulence in the presence of Couette typed mean flow but in the absence of rotation.
However, by experiment (e.g. 
\citealt{Paoletti_2012}), simulations in the context of accretion discs (e.g. \citealt{Avila_2012}) and in formation of 
large objects from the dusty gas surrounding a young star (e.g. \citealt{Cuzzi_2007, Ormel_2008}), transient growth in 
the case of otherwise linearly stable flows (e.g. \citealt{2005ApJ...629..383M, 2005ApJ...629..373A, Cantwell_2010, 
2011NJPh...13b3029M}), people argued for plausible emergence of hydrodynamic instability and hence further turbulence.
 
The idea of transient amplification (see \citealt{2002ApMRv..55B..57S},
for details) is quite popular to resolve similar issues in 
laboratory flows. Due to the presence of the large number of active nonnormal modes, the subcritical turbulence 
has quite rich, strongly nonlinear dynamics (see, e.g., homogeneous shear turbulence in a shearing 
box-like set up shown by \cite{Pumir_1996, Mamatsa_2016, Sekimoto_2016}). One of the most important
nonlinear processes in this case is the new fundamental cascade process, transverse cascade, which plays 
a key role in the self-sustaining dynamics of the turbulence. This further ensures regeneration of new 
transiently growing modes (\citealt{Mamatsa_2016}; see also \citealt{Gogicha_2017}, for MHD).
However, the Keplerian disc was questioned to have sustained purely hydrodynamic turbulence by this process
(e.g., \citealt{Lesur_2005}, also see \citealt{2005ApJ...629..383M}). Indeed, in direct numerical simulations at  
$Re \sim 10^5$, no sustained turbulence has been found (see, e.g.,  \citealt{Lesur_2005, Shen_2006_apj, 
Shi_2017}). Nevertheless, we believe that $Re\sim 10^5$ is still quite low for accretion discs to
rule out any hydrodynamic turbulence, where the Coriolis force is a strong hindering effect therein
to kill emergence of any instability and turbulence. We will demonstrate below that for a low $Re$,
the system should have been forced strongly to reveal instability.

We, therefore, search for a hydrodynamical origin of nonlinearity and hence plausible turbulence in the 
accretion disc.Our emphasis is the conventional linear instability when perturbation grows 
exponentially, unlike the case of transient growth.
We particularly consider here an extra force, to fulfill our purpose. \cite{2016ApJ...830...86N} initiated the study of hydrodynamics in the 
presence of an extra force in a simplistic model to observe the growth of perturbations in linear regime in astrophysical 
as well as in laboratory flows. The examples of the origin of such force in the context of biological sciences are: 
Brownian ratchets in soft condensed matter and biology (e.g. \citealt{Ait-Haddou2003, vanOudenaarden_1999, 
Parrondo_1996}), fluid-structure interaction in biological fluid dynamics (e.g. \citealt{peskin_2002}). However, in 
astrophysical context, particularly in accretion discs, the examples of origin of such force could be: the interaction 
between the dust grains and fluid parcel in protoplanetary discs (e.g. \citealt{Henning_1996}), back reactions of 
outflow/jet to accretion discs. These forces are also expected to be stochastic in nature. In fact,
much prior to that, \cite{Farrell_1993} explored the effect of stochastic force in the linearized Navier-Stokes 
equations. While it was already known that the maximal growth of threedimensional perturbation far
exceeds than that of twodimensional perturbation in channel flows, they wanted to check if stochastic
forcing further influences growth of perturbation. However, their exploration was limited to nonrotating
flows (or flows without Coriolis effect). Therefore, their results, while suggesting implications
to similar astrophysical flows as well, do not prove for it. This is important as astrophysical flows 
are generally involved with rotation and rotational (Coriolis) effect is prone to kill transient
amplification of perturbation (\citealt{2005ApJ...629..383M,2005ApJ...629..373A}).
Recently, the effect of stochastic forcing has been explored in the Keplerian
flow (\citealt{raz}) and it is shown based on the linear theory only that the 
zero mean stochastic forcing requires 
compressible fluid in order to transfer angular momentum in the shearing box 
approximation. However, the present work differs with respect to that of \cite{raz} in many aspects 
and it will be evident as we go along.
We explore rigorously the idea put forward by \cite{Farrell_1993} and \cite{2016ApJ...830...86N}, even generalizing it with nonlinear
effects. Due to the very presence of force (stochastic with nonzero mean or otherwise), we show in the present work that the amplitude of least stable perturbation 
for a Keplerian flow (and some laboratory flows as well) evolves to lead to nonlinearity and plausible turbulence in the 
system. This works even in incompressible fluid, but with the nonzero mean of 
stochastic force or finite (even if very small) effect of the force, if not
stochastic, in the flow equations. On the other hand, \cite{raz} considered 
multi-mode analysis to investigate transient growth of energy and transfer of angular momentum due to 
the perturbations in the presence of stochastic force.

The plan of the paper is the following. In \S\ref{sec:Landau equation in the presence of an extra force}, we establish 
the evolution of amplitude of the perturbations in a local disc in the presence of noise acting as an extra force. This 
is basically modified Landau equation describing nonlinear perturbation, in the presence of the Coriolis 
and external forces. Note that Landau equation in the context of accretion discs without extra force was explored by 
\cite{2011MNRAS.414..691R}. In \S\ref{sec:Results}, we present the results for perturbation evolution, along with its 
linear counterpart. By the eigenspectrum analysis in the linear regime, we
show how the extra force might affect the flow. We further discuss our results comparing them with the properties of conventional Landau equation 
(without Coriolis and extra forces) in \S\ref{diss}. We also argue therein, how the presence of force effectively 
changes the sign of growth rate, i.e. the least stable eigenvalue, 
based on the Landau
equation. We conclude in \S\ref{sec:Conclusion} that the 
presence of force makes the system nonlinear and hence reveals turbulence therein.


\section{Landau equation in the presence of an extra force and the Coriolis force}
\label{sec:Landau equation in the presence of an extra force}
We study the hydrodynamics at a local patch in the accretion disc. We assume 
the patch to be a cubical box of size $2L$. 
\begin{figure}
  \includegraphics[width = 8cm]{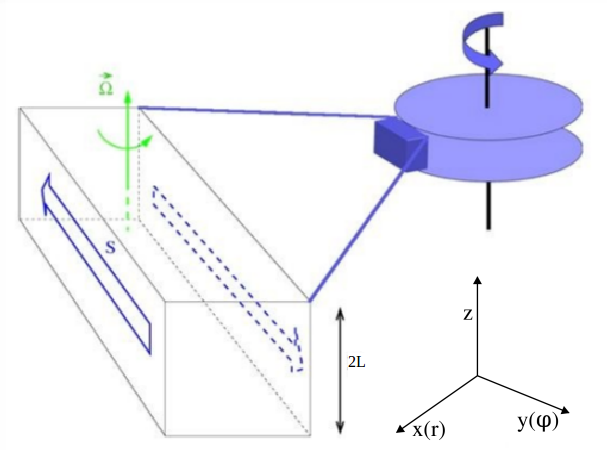}
  \caption{Model picture of local cubical box in accretion disc where we perform the analysis 
(http://ipag.obs.ujf-grenoble.fr/~longarep/astrophysics.html). Within the box, the Cartesian coordinate $x$ is 
along the radial cylindrical coordinate $r$ (with respect to the center of the accretion disc), $y$ is 
along $\phi$, and $z$ is same in both the systems.
}
  \label{fig:rotating_couette_flow}
 \end{figure}
The local flow geometry is shown in the Fig.~\ref{fig:rotating_couette_flow}, where we use Cartesian coordinates to 
describe the motion of the local fluid parcel. Usually, cylindrical polar coordinates are used to describe the dynamics 
of the flow in the accretion disc. However, the directions of the Cartesian coordinates within the box with 
respect to the cylindrical coordinates are shown in Fig.~\ref{fig:rotating_couette_flow}, where the Cartesian 
coordinates $x,y,z$ are along cylindrical coordinates $r,\phi,z$ respectively. The detailed description of the local 
formulation can be found in \citealt{2005ApJ...629..383M, MMR_2011}. As the fluid is in the local region, 
we assume the fluid to be incompressible as justified by \cite{2004A&A...425..385Y, 2005ApJ...629..383M, 
2005ApJ...629..373A, 2007A&A...463..817R, 2016ApJ...830...86N}. Here we recast the Navier-Stokes equation in 
Orr-Sommerfeld and Squire equations in the presence of additional stochastic force 
(\citealt{Farrell_1993}) and Coriolis force, by eliminating the 
pressure, utilizing the equation of continuity and ensemble averaging (see \citealt{2016ApJ...830...86N}), given by
\begin{eqnarray}
\begin{split}
 \left(\frac{\partial}{\partial t} + U\frac{\partial}{\partial y} \right)\nabla^2u - U''\frac{\partial u}{\partial y} 
+\frac{2}{q}\frac{\partial \zeta}{\partial z} \\- \frac{1}{Re}\nabla^4u + \Gamma_1 = NL^{u},
\label{eq:orr_sommerfeld_eq}
\end{split}
\end{eqnarray}

\begin{eqnarray}
\begin{split}
\left(\frac{\partial}{\partial t} + U\frac{\partial}{\partial y} \right)\zeta - U'\frac{\partial u}{\partial z} 
-\frac{2}{q}\frac{\partial u}{\partial z} \\- \frac{1}{Re}\nabla^2\zeta + \Gamma_2  = NL^{\zeta},
\label{eq:squire_eq}
\end{split}
\end{eqnarray}
where $u$ and $\zeta$ are respectively the $x$-component of the velocity and vorticity perturbations after ensemble 
averaging, $U$ the $y$-component of background velocity which for the present purpose of plane shear in the 
dimensionless units is $-x$ (see Appendix \ref{Modification of background flow in presence of force}), $q$ 
the rotation parameter with $\Omega(r)\propto 1/r^q$, $\Omega(r)$ being the angular frequency of the fluid parcel at 
radius $r$, $\Gamma$-s are the corresponding constant means of stochastic forces (white noise with nonzero mean due to 
gravity making the system biased, see \citealt{2016ApJ...830...86N})\footnote{This is equivalent to Brownian ratchet 
often proposed in biological systems. Here the net drift of Brownian motion is nonzero due to symmetry breaking 
effect.}\textsuperscript{,}\footnote{Let us say $X(t)$ be the random displacement variable of a Brownian motion 
with probability density function $P(X(t))$. Now, the stochastic time derivative of $X(t)$ will give the white 
noise. Due to, e.g., thermal fluctuation (however small it would be), the fluid parcel will do the random walk. However, 
due to the presence of gravity (for a Keplerian flow) or externally applied force (for plane Couette flow) there will be 
a preferential direction of the random walk and hence the random walk will be biased. Consequently, the white noise will 
have nonzero mean.} in the system described below and $NL$-s are the non-linear terms of perturbation. 
As described in Appendix \ref{Modification of background flow in presence of force}, in principle in the presence of 
force, background velocity should be modified with a quadratic variation of $x$ in the $y$-direction. However, depending 
on the force strength, the $x^2$-term may or may not be negligible with respect to the $x$-term. Indeed, for a very
small magnitude of this external force, $x^2$-term can be neglected keeping background velocity profile same as that 
without force, as shown explicitly in Appendix \ref{Modification of background flow in presence of force}. Also, the 
detailed derivation of equations (\ref{eq:orr_sommerfeld_eq}) and (\ref{eq:squire_eq}) is shown in Appendix 
\ref{sec:derivation of OS and Sqiuire eqs}. The $x$-component of vorticity and the non-linear terms are given by
\begin{eqnarray}
 \zeta &=& \frac{\partial w}{\partial y} - \frac{\partial v}{\partial z},
 \label{eq:x_comp_vorti}
 \\
 NL^u &=& -\nabla^2\{(\textbf{u}'\cdot \nabla)u\}+\frac{\partial}{\partial x} \nabla \cdot \{(\textbf{u}'\cdot \nabla) 
\textbf{u}'\},
 \label{eq:non_orr_sommerfeld_eq}
 \\
 NL^{\zeta} &=& -\frac{\partial}{\partial y}\{(\textbf{u}'\cdot \nabla)w\} + \frac{\partial}{\partial 
z}\{(\textbf{u}'\cdot \nabla) v\},
 \label{eq:non_squire_eq}
\end{eqnarray}
where $\textbf{u}'=(u,v,w)$, which is the perturbed velocity vector, the derivation of 
equations (\ref{eq:non_orr_sommerfeld_eq}) and (\ref{eq:non_squire_eq}) is also shown in Appendix 
\ref{sec:derivation of OS and Sqiuire eqs}. However, \cite{Farrell_1993} assumed that the perturbation
itself is stochastic without considering possible change in background flow due the forcing. 
They argued that the stochasticity in the dynamical system stems from the random nature of 
the forcing arisen  during perturbation, in our case $\Gamma_{1,2}$, more precisely their properties before ensemble averaging,
i.e. $F_{1,2}$ or $F_{x,y,z}$, as shown in Appendix \ref{sec:derivation of OS and Sqiuire eqs}.

Note that the flow variables,
$u$ and $\zeta$, become stochastic variables due to the effect of stochastic force in the flow. 
Hence, we ensemble average this stochasticity while we derive the temporal dependence of the perturbation in linear and 
nonlinear regimes.
The linearized versions of equations (\ref{eq:orr_sommerfeld_eq}) and (\ref{eq:squire_eq}) before ensemble averaging are 
given in equation (1) in \citealt{Farrell_1993} and equations 
(1) and (2) in \citealt{2016ApJ...830...86N} and they also can be obtained from equations 
(\ref{eq:orr_sommerfeld_eq_before_ensm_avg}) and 
(\ref{eq:squire_eq_before_ensm_avg}) in Appendix \ref{sec:derivation of OS and Sqiuire eqs} by removing the nonlinear 
terms. Equations 
(\ref{eq:orr_sommerfeld_eq}) and 
(\ref{eq:squire_eq}), along with the equation of continuity for incompressible flow given by
\begin{equation}
 \nabla \cdot \textbf{u}' = 0,
 \label{eq:continuity_eq}
\end{equation}
form the solvable system of differential equations. We choose the no-slip boundary conditions along $x$ 
direction (\citealt{1970JFM....40...97E, 2004A&A...425..385Y, 2005ApJ...629..383M, 2007A&A...463..817R}), i.e. $u = 
v = w = 0$ at $x=\pm1$ or equivalently
\begin{equation}
 u = \frac{\partial u}{\partial x} = \zeta = 0,\,\ {\rm at}\,\, x = \pm1.
 \label{eq:no-slip_bc}
\end{equation}
However, we consider periodic boundary conditions in $y$ and $z$ directions, as the perturbations in these directions can 
be 
written in terms of Fourier modes due to the translational invariance of the background flow along these directions. It 
is well 
known (e.g. \citealt{Lin_1961, Butler&Farrell_1992, 2005ApJ...629..383M}) that the solutions for the homogeneous 
part of equations (\ref{eq:orr_sommerfeld_eq}) and (\ref{eq:squire_eq}) with nonzero viscosity will form a complete set 
of discrete 
eigenmodes. However interestingly note that earlier \cite{2005ApJ...629..383M} and \cite{2005ApJ...629..373A} 
showed the solutions of Orr-Sommerfeld and Squire equations in the context of linear instability 
in accretion discs practically do not depend on the fact whether $x$ is bounded or extended in
infinite domain.

\subsection{Plausible source of extra force}
\label{subsec:source_of_force}
We propose two plausible sources for the force in the context of accretion disc. One could be due to the 
dust-grain in protoplanetary disc interacting with the fluid flow and the other one could be the feedback from jet or 
outflow onto the 
accretion disc. These two processes could be modeled considering fluid-particle interactions 
(\citealt{doi:10.1080/03605300500394389}). 
Let us assume that $f(\textbf{r}, \bm{\xi}, t)d^3\xi$ be the number per unit volume of spherical 
particles of radius $a$ at position $\textbf{r}$, having 
velocity within $\bm{\xi}$ and $\bm{\xi}+d\bm{\xi}$, which may describe the grains floating in the protoplanetary disc. 
The 
force, therefore, on a particle by the fluid parcel is $6\pi\mu a(\bm{\xi}-\textbf{U})$, where 
$\mu$ is the dynamical viscosity and $\textbf{U}$ is the fluid velocity. On the other hand, the force acting on the fluid 
parcel of unit mass by the particles is $\int_{\bm{\xi}}{6\pi\nu 
a(\textbf{U}-\bm{\xi})fd\bm{\xi}}$, where $\nu$ is the kinematic viscosity of the fluid of density 
$\rho$ and is defined by $\mu/\rho$. Now, the number density function $f$ is expected to be stochastic in nature for 
both the cases in the context of accretion discs due to the stochastic nature of motion of floating dust-grains and 
feedback, hence the force is. 
Let us consider the velocity of the particles has radial dependence, i.e. $\xi = \xi(\textbf{r})$. Since the analysis is 
done in a 
shearing box at a particular radius with a very small radial width, we 
assume the number of particles per unit volume within the shearing box 
be $f(t)\Delta^3\xi$. The force acting on the fluid parcel of unit mass by the particles, therefore, is $6\pi \nu a 
(\textbf{U} - 
\bm{\xi}) f(t)\Delta^3\xi$. As described in Appendix \ref{sec:derivation of OS and Sqiuire eqs} in detail, particularly 
in equation
(\ref{eq:N-S_eqn_dim_less}), we can consider the background stochastic force to
be $\textbf{F}'' \cong 6\pi \nu a (\textbf{U} - 
\bm{\xi}) f(t)\Delta^3\xi$. If we perturb the flow, $\textbf{U}$ will be replaced by $\textbf{U}+\textbf{u}'$ and 
$\textbf{F}''$ will become 
$6\pi \nu a (\textbf{U} - \bm{\xi}) f(t)\Delta^3\xi + 6\pi \nu a \textbf{u}' f(t)\Delta^3\xi$. After the background 
subtraction, the extra force, $\textbf{F}$, becomes $6\pi \nu a \textbf{u}' f(t)\Delta^3\xi$, when at a particular 
radius, $f(t)\Delta^3\xi$ appears to be 
independent of spatial coordinates. 
According to \cite{Farrell_1993} however, any forcing arises due to perturbation only. Hence,
there is no change of background velocity and above force $\textbf{F}=6\pi \nu a \textbf{u}' f(t)\Delta^3\xi$
directly impacts in the system during perturbation only and any such forcing arises after 
background subtraction.
In either of the cases, as described in Appendix \ref{sec:derivation of OS and Sqiuire eqs}, particularly in 
equations 
(\ref{eq:F_1}) and (\ref{eq:F_2}), the components of extra force are therefore
\begin{eqnarray}
 F_1 &=& \mathcal{K}\nabla^2 u,
     \label{eq:astro_F_1_final}\\
 F_2  &=& \mathcal{K} \zeta,
     \label{eq:astro_F_2_final}
\end{eqnarray}
where $\mathcal{K}$ is $6\pi \nu a f(t)\Delta^3\xi$.

Apparently the extra force is then involved with the solution itself.
Hence in principle, in the context of the said model, $F_1$ and $F_2$ can 
be combined with the corresponding first term of equations 
(\ref{eq:orr_sommerfeld_eq}) and (\ref{eq:squire_eq}) respectively. Subsequently,
depending on $\mathcal{K}$, stability of flow may be influenced compared to the
case without forcing. However, due to the very stochastic nature of the force, 
equations (\ref{eq:astro_F_1_final}) and (\ref{eq:astro_F_2_final}) turn out to
be stochastic in nature, hence they have to be ensemble averaged in order to 
determine the temporal dependence of the perturbation. Nevertheless, unlike other
terms in equations (\ref{eq:orr_sommerfeld_eq}) and (\ref{eq:squire_eq}), $u$ and
$\zeta$ cannot be trivially separated out from $\mathcal{K}$ while ensemble 
averaging $\mathcal{K}\nabla^2 u$ and $\mathcal{K} \zeta$. Hence, for the present
purpose, we a priori assume them to be $\Gamma_1$ and $\Gamma_2$. Indeed,
for any other force model, e.g. thermal fluctuation in fluid elements (which is quite a
common choice in statistical and condensed matter systems),
$F_1$ and $F_2$ could be quite different and need to be modeled separately.
Hence, for generic purpose also, $\Gamma_1$ and $\Gamma_2$ are chosen to be 
constant a priori for the present purpose. We assume any time-dependences,
even if arisen from $u$ and $\zeta$, averaged out due to their association
with random number $\mathcal{K}$.

Now for micrometer size grains and width of shearing
box of $0.1$ Schwarzschild radius, around a $m$ solar mass central object
$\mathcal{K}\sim 2\times 10^{6}\times m^2 f^\prime(t)\Delta^3\xi^\prime/Re$,
where quantities with ``prime" denote their dimensionful values. Obviously,
larger $Re$ corresponds to smaller force, which is at per expectation. Similar
scaling is true for laboratory flows. For a protoplanetary disc around 
a 10 solar mass central object with number density of grain $\sim 10^{11}$ cm$^{-3}$
(when a typical midplane total number density $\sim 10^{13}$ cm$^{-3}$),
$\mathcal{K}\sim 2\times 10^5$ for $Re\sim 10^{14}$ (see, e.g., \citealt{2013PhLB..721..151M}, for bounds on disc $Re$).

Had the force not been stochastic in nature or flow 
variables been separated out from $\mathcal{K}$ even after ensemble
averaging, then a linear stability analysis could be performed
for the linearized set of equations (\ref{eq:orr_sommerfeld_eq}) and 
(\ref{eq:squire_eq}) in the same spirit of, e.g., \cite{2005ApJ...629..383M} except
with modified coefficients of $\nabla^2u$ and $\zeta$. 
This effect has been explored in \S \ref{sec:Results} with examples.

Such forcing has already been demonstrated in biological systems with incompressible fluid (\citealt{peskin_2002}). Apart from this, \cite{Ioannou_2001} mentioned that stochastic forcing in the context of
accretion discs could be due to nonlinear terms which are otherwise neglected because of linearisation or due to external processes
such as tidal interaction in binaries, outbursts in binary systems, or perturbation debris from shock waves. 
Note that very tiny thermal fluctuation in fluids may lead to stochastic motion, however small be, of particles. See Appendix 
\ref{sec:derivation of OS and Sqiuire eqs} for survival of such force after ensemble averaging. See also 
\cite{2016ApJ...830...86N} and 
references therein, describing other plausible origin of force.

\subsection{Linear Theory}
In the evolution of linear perturbation,
let the linear solutions be
\begin{eqnarray}
 u = \hat u(x,t) e^{i\textbf{k}\cdot \textbf{r}},\\ 
 \zeta = \hat{\zeta}(x,t) e^{i\textbf{k}\cdot \textbf{r}},
\end{eqnarray}
with $\textbf{k} = (0, k_y, k_z)\ {\rm and}\ \textbf{r} = (0, y, z).$
Substitute these in equations (\ref{eq:orr_sommerfeld_eq}) and (\ref{eq:squire_eq}), neglecting non-linear terms, we 
obtain
\begin{equation}
\begin{split}
 (\mathcal{D}^2 - k^2)\frac{\partial \hat {u}}{\partial t} + ik_yU(\mathcal{D}^2 - k^2)\hat{u} - 
U''ik_y\hat{u}+\frac{2}{q}ik_z\hat{\zeta}\\-\frac{1}{Re}(\mathcal{D}^2 - k^2)^2\hat{u}+\Gamma_1 e^{-i\textbf{k}\cdot 
\textbf{r}} =0
\label{eq:linear_orr_sommerfeld_eq}
\end{split}
\end{equation}
and
\begin{equation}
 \begin{split}
  \frac{\partial \hat {\zeta}}{\partial t} 
+ik_yU\hat{\zeta}-\left(U'+\frac{2}{q}\right)ik_z\hat{u}-\frac{1}{Re}(\mathcal{D}^2 - 
k^2)\hat{\zeta}\\+\Gamma_2 e^{-i\textbf{k}\cdot \textbf{r}} =0,
\label{eq:linear_squire_eq}
 \end{split}
\end{equation}
where $\mathcal{D} = \frac{\partial}{\partial x}$.
Recasting equation (\ref{eq:linear_orr_sommerfeld_eq}) we obtain
\begin{eqnarray}
 \begin{split}
  \frac{\partial \hat {u}}{\partial t} + i(\mathcal{D}^2 - k^2)^{-1}\Bigl[k_yU\left(\mathcal{D}^2-k^2\right) - 
k_yU''\\-\frac{1}{iRe}(\mathcal{D}^2 - k^2)^2\Bigr]\hat{u} +(\mathcal{D}^2 - k^2)^{-1}\frac{2}{q}ik_z\hat{\zeta} 
\\+(\mathcal{D}^2 - 
k^2)^{-1}\Gamma_1 e^{-i\textbf{k}\cdot \textbf{r}} = 0.
\label{eq:linear_orr_sommerfeld_eq_recast}
 \end{split}
\end{eqnarray}

Further combining equations (\ref{eq:linear_orr_sommerfeld_eq_recast}) and (\ref{eq:linear_squire_eq}) we obtain 
\begin{eqnarray}
 \begin{split}
  \frac{\partial}{\partial t}Q +i\mathcal{L}Q +\Gamma = 0,
  \label{eq:lin_Q_eq}
 \end{split}
\end{eqnarray}
where
\begin{eqnarray}
\begin{split}
 Q = \begin{pmatrix}\hat{u}\\ \hat{\zeta}\end{pmatrix}, \ 
 \mathcal{L} = \begin{pmatrix}
                \mathcal{L}_{11}& \mathcal{L}_{12}\\
                \mathcal{L}_{21}& \mathcal{L}_{22}
               \end{pmatrix},
\label{eq:the_L_matrix}
\end{split}
\end{eqnarray}

\begin{eqnarray*}
\begin{split}
 \mathcal{L}_{11} =& (\mathcal{D}^2 - 
k^2)^{-1}\Bigl[k_yU\left(\mathcal{D}^2-k^2\right)-k_yU''\\&-\frac{1}{iRe}\left(\mathcal{D}^2-k^2\right)^2\Bigr],\\
 \mathcal{L}_{12} =& \left(\mathcal{D}^2-k^2\right)^{-1}\frac{2k_z}{q},\\
 \mathcal{L}_{21} =& -\left(U'+\frac{2}{q}\right)k_z,\\
 \mathcal{L}_{22} =& k_yU-\frac{1}{iRe}\left(\mathcal{D}^2-k^2\right),
\end{split}
\end{eqnarray*}
and
\begin{equation}
 \Gamma = e^{-i\textbf{k}\cdot \textbf{r}}\begin{pmatrix}
           (\mathcal{D}^2 - k^2)^{-1}\Gamma_1\\
           \Gamma_2
          \end{pmatrix}.
          \label{eq:Gamma}
\end{equation}
Let us subsequently assume the trial solution of equation (\ref{eq:lin_Q_eq}) be
\begin{equation}
 Q = AQ_xe^{-i\sigma t}-\frac{1}{\mathcal{D}_t+i\mathcal{L}}\Gamma,
 \label{eq:lin_trial_sol}
\end{equation}
where $\sigma$ is the eigenvalue corresponding to the particular mode and it is complex having real ($\sigma_r$) and 
imaginary ($\sigma_i$) parts,
 \begin{equation}
  Q_x = \begin{pmatrix}\phi^u(x)\\ \phi^{\zeta}(x)\end{pmatrix}
  \label{eq:Q_x}
 \end{equation}
and $\mathcal{D}_t$ stands for ${\partial}/{\partial t}.$ 
$Q_x$ is the eigenfunction corresponding to the homogeneous part of equation (\ref{eq:lin_Q_eq}), i.e. $Q_x$ 
satisfies $\mathcal{L}Q_x = \sigma Q_x$. The first term of right hand side of equation (\ref{eq:lin_trial_sol}) is due to 
the homogeneous part of equation (\ref{eq:lin_Q_eq}) and the second term is due to the inhomogeneous part, i.e. the 
presence of 
$\Gamma$, of the same equation.
Hence, $Q$ is influenced by the force $\Gamma$.

\subsection{Non-linear theory}
For the non-linear solution, following similar work but in the absence of 
force, e.g. \citealt{1970JFM....40...97E,schmid2001stability,2002ApMRv..55B..57S,2011MNRAS.414..691R}, we assume the series solution for 
velocity and 
vorticity, i.e. 
\begin{eqnarray}
 u = \sum_{n\rightarrow -\infty}^{\infty} u_n = \sum_{n\rightarrow -\infty}^{\infty} \bar{u}_n(t,x)e^{in(\textbf{k}\cdot 
\textbf{r} - 
\sigma_r t)},\\
 \zeta = \sum_{n\rightarrow -\infty}^{\infty} \zeta_n = \sum_{n\rightarrow -\infty}^{\infty} \bar{\zeta}_n(t,x) 
e^{in(\textbf{k}\cdot 
\textbf{r} - \sigma_r t)},
\end{eqnarray}
when obviously $\bar{u}_{-n} = \bar{u}^*_n$ and $\bar{\zeta}_{-n}=\bar{\zeta}^*_n$. 
This approach will help in comparing our solutions in accretion discs with the existing literature,
without losing any important physics, as will be evident below.

We substitute these in equations (\ref{eq:orr_sommerfeld_eq}) and (\ref{eq:squire_eq}) and obtain
\begin{eqnarray}
 \begin{split}
  \sum_{n=-\infty}^{+\infty}\Bigl[\Bigl\{\left(\mathcal{D}^2-n^2k^2\right)\frac{\partial}{\partial t} - 
in\sigma_r\left(\mathcal{D}^2-n^2k^2\right)\\+ 
ink_yU\left(\mathcal{D}^2-n^2k^2\right)-U''ink_y\Bigr\}\bar{u}_n+\frac{2}{q}nik_z\bar{\zeta}_n\\-\frac{1}{Re}
\left(\mathcal{D}
^2-n^2k^2\right)^2\bar{u}_n\Bigr ]
e^{in(\textbf{k} \cdot\textbf{r}-\sigma_r t)} +\Gamma_1 \\= NL^u_n e^{in(\textbf{k}\cdot\textbf{r}-\sigma_r t)}
 \end{split}
\end{eqnarray}
and 
\begin{eqnarray}
 \begin{split}  
\sum_{n=-\infty}^{+\infty}\Bigl[\Bigl\{\frac{\partial}{\partial 
t}-in\sigma_r+Uink_y-\frac{1}{Re}\left(\mathcal{D}^2-n^2k^2\right)\Bigr\}
\bar{\zeta}_n \\-\left(U'+\frac{2}{q}\right)ink_z\bar{u}_n\Bigr] e^{in(\textbf{k}\cdot\textbf{r}-\sigma_r t)} +\Gamma_2 
\\=NL^{\zeta}_n 
e^{in(\textbf{k}\cdot\textbf{r}-\sigma_r t)}.
 \end{split}
\end{eqnarray}
Now, we collect the coefficients of the term $e^{i(\textbf{k}\cdot\textbf{r}-\sigma_r t)}$,
to capture least nonlinear effect following, e.g., \cite{1970JFM....40...97E, 2011MNRAS.414..691R}, from both sides and 
obtain

\begin{eqnarray}
 \begin{split}
 \frac{\partial \bar{u}_1}{\partial t}-i\sigma_r 
\bar{u}_1+i\Bigl[\left(\mathcal{D}^2-k^2\right)^{-1}\Bigl\{k_yU\left(\mathcal{D}
^2-k^2\right)\\-U''k_y-\frac{1}{Re}\left(\mathcal{D}^2-k^2\right)^2\Bigr\}\Bigr]\bar{u}_1\\+\frac{2}{q}ik_z\left(\mathcal{
D}^2-k^2\right)^{-1}
\bar{\zeta}_1= \left(\mathcal{D}^2-k^2\right)^{-1}NL^{u}_1
\label{eq:1st_nonlin_orr_sommerfeld_eq}
\end{split}
\end{eqnarray}
and 
\begin{eqnarray}
 \begin{split}  
  \frac{\partial \bar{\zeta}_1}{\partial t}-i\sigma_r \bar{\zeta}_1+Uik_y 
  \bar{\zeta}_1-\frac{1}{Re}\left(\mathcal{D}^2-k^2\right)\bar{\zeta}_1\\-\left(U'+\frac{2}{q}\right)ik_z\bar{u}_1 
=NL^{\zeta}_1.
  \label{eq:1st_nonlin_squire_eq}
 \end{split}
\end{eqnarray}
Note that $NL^{u}_n$ and $NL^{\zeta}_n$ contain various combinations of $e^{i(\textbf{k}\cdot\textbf{r}-\sigma_r t)}$. 
See Appendix 
\ref{The nonlinear terms for three dimensional perturbations} for details.
If we assume further \begin{equation}
 Q_1 = \begin{pmatrix}\bar{u}_1(x,t)\\ \bar{\zeta}_1(x,t)\end{pmatrix},
  \label{eq:Q_1}
\end{equation}
we can combine equations (\ref{eq:1st_nonlin_orr_sommerfeld_eq}) and (\ref{eq:1st_nonlin_squire_eq}) to obtain
\begin{equation}
 \frac{\partial Q_1}{\partial t} -i\sigma_r Q_1 + i\mathcal{L}Q_1 = NL_1,
 \label{eq:Q_1_nonlin_eq_time_evo}
\end{equation}
where $NL_1 = \begin{pmatrix}\left(\mathcal{D}^2-k^2\right)^{-1}NL^{u}_1\\ NL^{\zeta}_1\end{pmatrix}$.
We assume the solution for $Q_1$ to be
\begin{equation}
 Q_1 = \sum_{m=1}^{\infty}{A_{t,m}} {Q_{x,m}} - \frac{1}{\mathcal{D}_t+i\mathcal{L}}\Gamma,
 \label{eq:Q_1_total_trial_sol}
\end{equation}
where $m$ stands for various eigenmodes.

However, to the first approximation, our interest is in the least stable mode. See \citealt{1970JFM....40...97E} for 
similar description in 
two dimensions without $\Gamma$ and \citealt{2011MNRAS.414..691R} for three dimensional Keplerian disc 
without $\Gamma$. We, therefore, omit the summation and subscript $m$ in equation (\ref{eq:Q_1_total_trial_sol}) and 
obtain
\begin{equation}
 Q_1 = {A_{t}} {Q_{x}} - \frac{1}{\mathcal{D}_t+i\mathcal{L}}\Gamma.
 \label{eq:Q_1_trial_sol}
\end{equation}
We then substitute equation (\ref{eq:Q_1_trial_sol}) in equation (\ref{eq:Q_1_nonlin_eq_time_evo}) and obtain
\begin{eqnarray*}
\begin{split}
&Q_x\frac{dA_t}{dt}-\frac{\partial}{\partial t}\Bigl(\frac{1}{\mathcal{D}_t+i\mathcal{L}}\Bigr)\Gamma - i\sigma_rA_tQ_x 
&\\ &+i\sigma_r\Bigl(\frac{1}{\mathcal{D}_t+i\mathcal{L}}\Bigl)\Gamma +iA_t\mathcal{L}Q_x - 
i\mathcal{L}\Bigl(\frac{1}{\mathcal{D}_t+i\mathcal{L}}\Bigl)\Gamma& = NL_1
\end{split}
\end{eqnarray*}

\begin{eqnarray*}
 \begin{split}
  \Rightarrow &Q_x\frac{dA_t}{dt} - (\mathcal{D}_t+i\mathcal{L})\Bigl(\frac{1}{\mathcal{D}_t+i\mathcal{L}}\Bigr)\Gamma + 
A_t\underbrace{(-i\sigma Q_x+i\mathcal{L}Q_x)}_\text{$0$}&\\ &-\sigma_i A_t Q_x +i\sigma_r 
\Bigl(\frac{1}{\mathcal{D}_t+i\mathcal{L}}\Bigr)\Gamma = |A_t|^2A_t\mathcal{S}&
 \end{split}
\end{eqnarray*}

\begin{eqnarray*}
\begin{split}
\Rightarrow Q_x\frac{dA_t}{dt} 
-\sigma_iA_tQ_x \underbrace{-\Gamma+i\sigma_r\Bigl(\frac{1}{\mathcal{D}_t+i\mathcal{L}}\Bigr)\Gamma}_\text{$\Gamma'$} = 
|A_t|^2A_t\mathcal{S}
\label{eq:prior to landau eq}
\end{split}
\end{eqnarray*}

\begin{eqnarray}
\Rightarrow Q_x\frac{dA_t}{dt}-\sigma_iA_tQ_x + \Gamma' = |A_t|^2A_t\mathcal{S},
\end{eqnarray}
where the detailed calculation for $\Gamma'$ is shown in the Appendix \ref{sec:The calculation of Gamma}. 
$\mathcal{S}$ is the spatial contribution from nonlinear term,
computed following \cite{2011MNRAS.414..691R}\footnote{We consider a slightly different notation for nonlinear terms. 
We keep number $n$ as a subscript, while \cite{2011MNRAS.414..691R} used it as a superscript, e.g. 
we use $NL^{u}_1$ and \cite{2011MNRAS.414..691R} used $NL^{u1}$.}
where our notation $\mathcal{S}$ is represented as $\begin{pmatrix}\mathcal{S}u \\ \mathcal{S}\zeta\end{pmatrix}$. Please note the section 2.4.1, Appendix B and Appendix C
of \cite{2011MNRAS.414..691R} to have the details of $\mathcal{S}$.
To obtain $\mathcal{S}$ using separation of variables and for sufficiently small and slowly varying amplitude, we assume 
the 
following:
\begin{enumerate}
 \item $A_t$ is so small that $\dot{A_t}/A_t$ is approximately $\sigma_i$.
 \item $\frac{\partial}{\partial x}\Bigl(\frac{1}{\mathcal{D}_t+i\mathcal{L}}\Bigr)\Gamma$ is negligible compared to 
$\frac{\partial}{\partial t}\Bigl(\frac{1}{\mathcal{D}_t+i\mathcal{L}}\Bigr)\Gamma$ as 
$||\mathcal{D}_t^2||>||\mathcal{L}^2||.$
\end{enumerate}

This is similar to what was considered by \cite{1970JFM....40...97E} and \cite{2011MNRAS.414..691R}. Now we utilize the 
bi-orthonormality between $Q_x$ and its conjugate function $\tilde{Q}_x$ and from equation 
(\ref{eq:prior to landau 
eq}) we obtain
\begin{eqnarray}
\frac{dA_t}{dt} - \sigma_i A_t +\mathcal{N} = p|A_t|^2A_t, 
\label{eq:mod_landau_eq}
\end{eqnarray}
where
\begin{eqnarray}
\mathcal{N} = \int_{-1}^{1}{dx \tilde{Q}_x^{\dagger}\Gamma'}
\label{eq:N}
\end{eqnarray}
and 
\begin{eqnarray}
p = \int_{-1}^{1}{dx \tilde{Q}_x^{\dagger}\mathcal{S}}.
\end{eqnarray}
Again, we recall the expression for $\Gamma'$ as
\begin{eqnarray}
\begin{split}
\Gamma' =&-\Gamma + i\sigma_r\Bigl(\frac{1}{\mathcal{D}_t+i\mathcal{L}}\Bigr)\Gamma& \\&= - \Gamma + 
i\sigma_r(t-i\mathcal{L}t^2)(1+\mathcal{L}^2t^2)^{-1}\Gamma.&
\label{eq:exp_for_Gamma'}
\end{split}
\end{eqnarray}

Throughout the paper, $\Gamma$ from equation (\ref{eq:Gamma}) has been decomposed as  
$\Gamma \rightarrow \Gamma \begin{pmatrix}
          1\\
           1
          \end{pmatrix}$ by adjusting $\Gamma_1$ and $\Gamma_2$, as they are only the free parameters.
          

\section{Evolution of perturbations}
\label{sec:Results}

\begin{figure}
\includegraphics[width=\columnwidth]{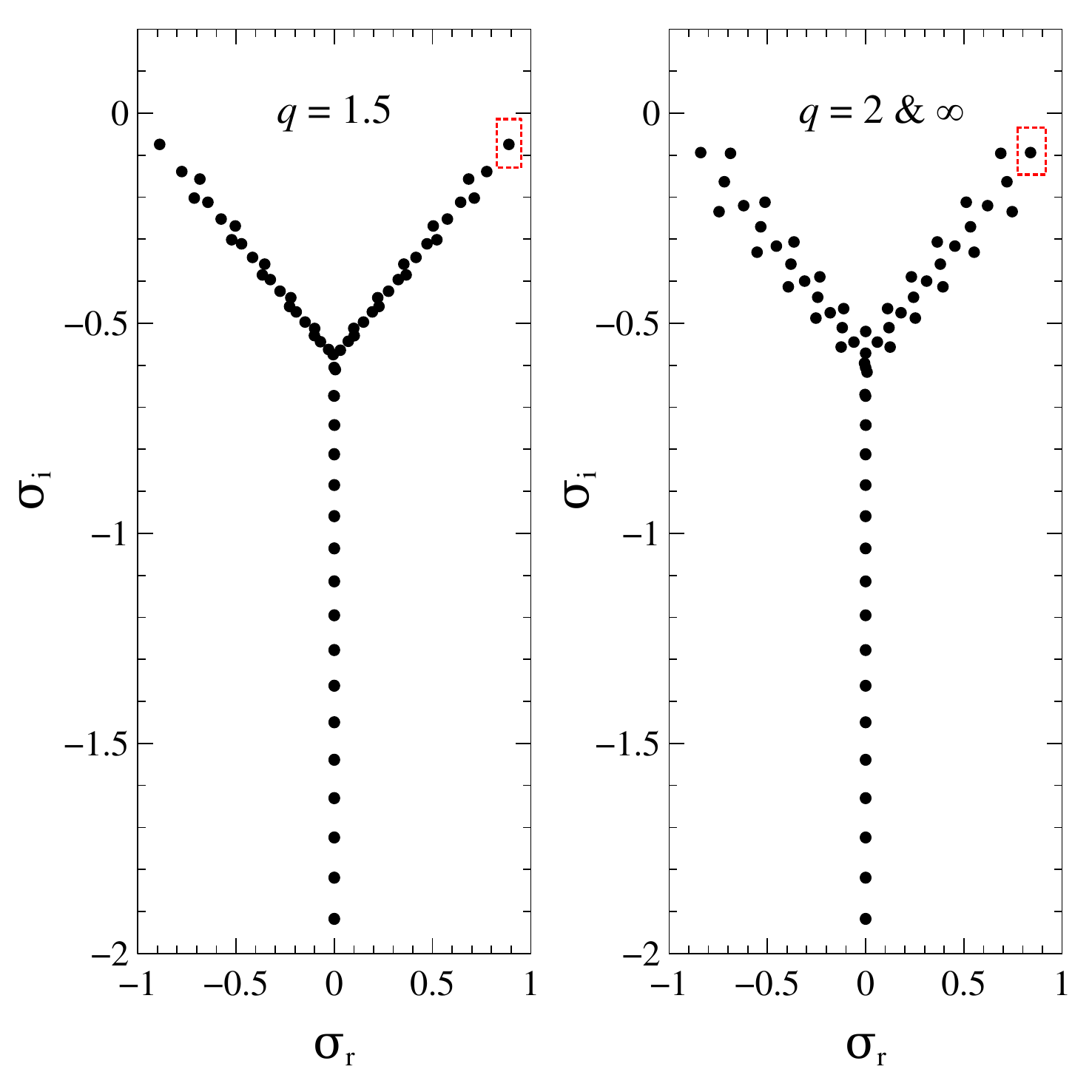}
\caption{Variation of $\sigma_i$ with $\sigma_r$ for $Re = 2000$ and $k_y=k_z = 1$ for the Keplerian flow ($q=1.5$), 
constant 
angular momentum flow ($q=2$) and plane Couette flow ($q\rightarrow\infty$). The latter two 
eigenspectra are identical. The dotted box represents the least stable mode for the 
respective cases.}
\label{fig:eval_three_cases}
\end{figure}

\begin{figure}
\includegraphics[width=\columnwidth]{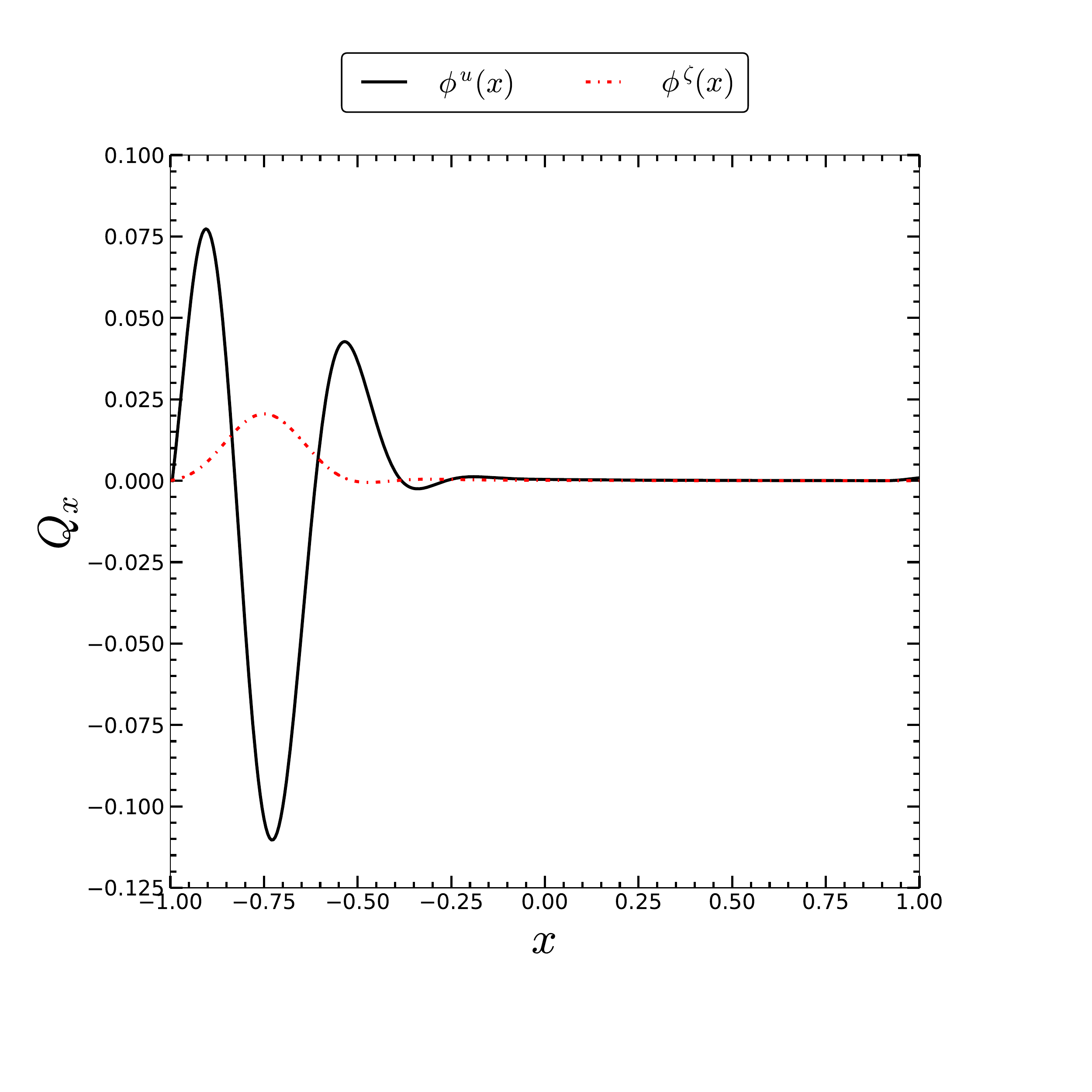}
\caption{Variation of eigenfunction ($Q_x$) for the least stable mode as a function of $x$ for $Re = 1000$ and $k_y=k_z = 
1$ for the Keplerian 
flow ($q=1.5$).}
\label{fig:evec_OS_Squire_comb_3d_kep_R_1000_N_399_ky_1_kz_1_mod}
\end{figure}

We explore here the evolution of the perturbation amplitude based on equation (\ref{eq:mod_landau_eq}). Note, equation 
(\ref{eq:mod_landau_eq}) is a nonlinear equation. Nevertheless, we explore the results for the linear and nonlinear 
evolutions both, when 
for the former, we neglect R.H.S. of equation (\ref{eq:mod_landau_eq}). However, the typical eigenspectra, for linearized 
Keplerian flow 
($q=1.5$), constant angular momentum flow ($q=2$) and plane Couette flow ($q\rightarrow\infty$), for $Re = 2000$ and 
$k_y=k_z=1$ are shown 
in the Fig.~\ref{fig:eval_three_cases}. $\mathcal{L}_{12}$ and $\mathcal{L}_{21}$ in equation (\ref{eq:the_L_matrix}) are 
zero for the 
plane Couette flow and constant angular momentum flow respectively. This is the reason for obtaining same eigenspectra 
for both plane 
Couette and constant angular momentum flows. We perform the whole analysis for the least stable modes for the respective 
flows and these 
least stable modes are shown in dotted box in Fig.~\ref{fig:eval_three_cases}.
A representative sample eigenvector is displayed in Fig.~\ref{fig:evec_OS_Squire_comb_3d_kep_R_1000_N_399_ky_1_kz_1_mod}.

\begin{figure}
\includegraphics[width=\columnwidth]{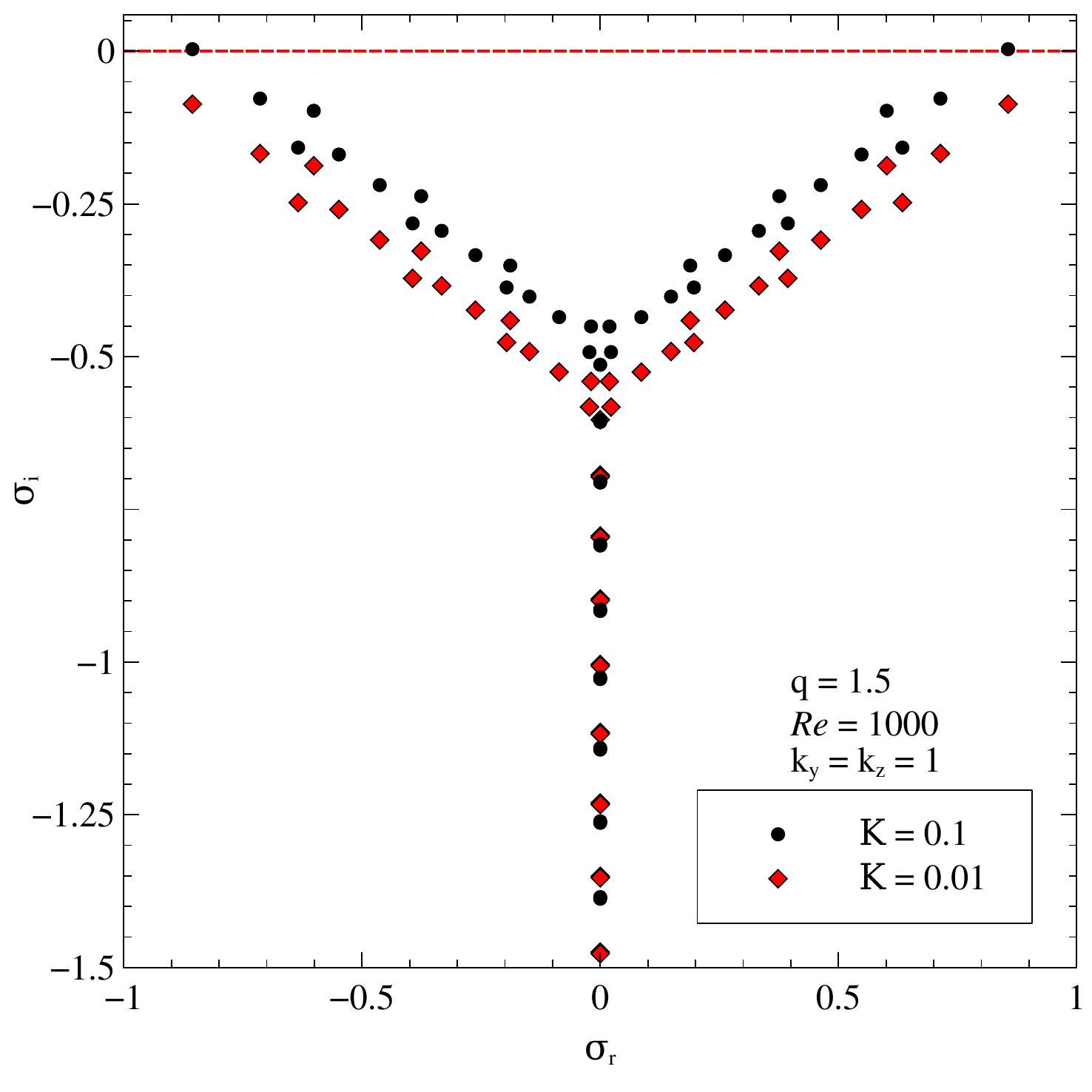}
\caption{Eigenspectra for the Keplerian flow with $\mathcal{K} = 0.1$ and $0.01$, for $Re = 1000$ 
and $k_y = k_z = 1$. Note the uppermost two eigenvalues for $\mathcal{K}=0.1$ with positive
$\sigma_i$.}
\label{fig:eval_OS_Squire_3d_mod_kep_1000_diff_kappa}
\end{figure}

\begin{figure}
\includegraphics[width=\columnwidth]{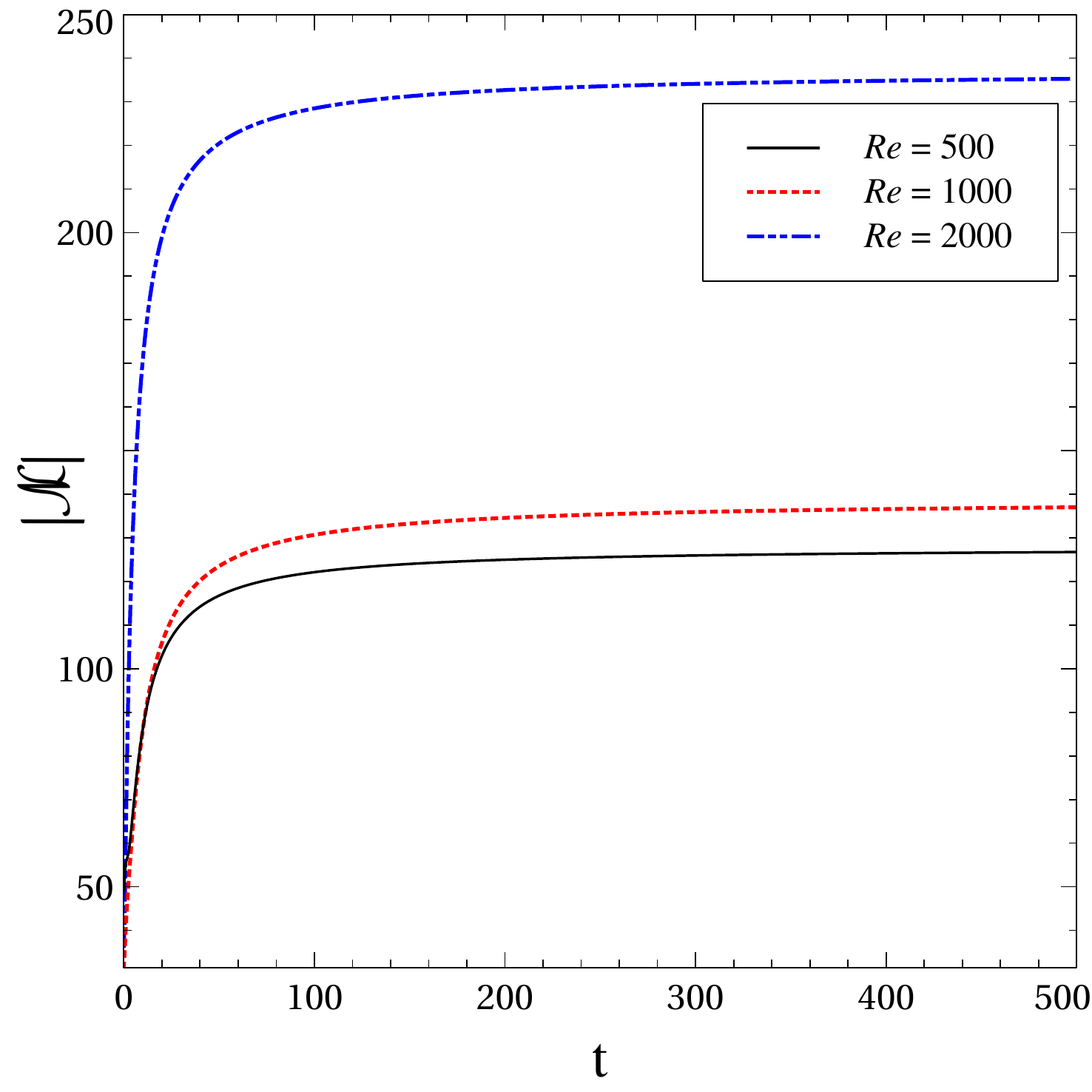}
\caption{Variation of $\mathcal{N}$ as a function of $t$ for $Re=500, \ 1000$ and 2000, for $\Gamma = 10^4$ and 
$k_y=k_z=1$, 
corresponding to the respective least stable modes.}
\label{fig:N_vs_t_R_500_1000_2000}
\end{figure}

As described in \S 2.1, particularly for equations (\ref{eq:astro_F_1_final}) 
and (\ref{eq:astro_F_2_final}), if 
the external force is, e.g., not stochastic in nature, then the effect of force can easily
be encoded in the coefficients of $\nabla^2 u$ and $\zeta$ in equations (\ref{eq:orr_sommerfeld_eq}) and 
(\ref{eq:squire_eq}) respectively, where $\Gamma_1=F_1$ and $\Gamma_2=F_2$. 
Fig.~\ref{fig:eval_OS_Squire_3d_mod_kep_1000_diff_kappa} describes the eigenspectra for the Keplerian flow with $Re = 
1000$ 
for $\mathcal{K} = 0.1$ and $0.01$ for the linearized set of equations 
(\ref{eq:orr_sommerfeld_eq}) and (\ref{eq:squire_eq}) (i.e. $NL^u=NL^\zeta=0$) and 
$\Gamma_1=F_1$ and $\Gamma_2=F_2$. While $\mathcal{K} = 0.1$ makes the flow 
unstable, $\mathcal{K} = 0.01$ cannot. This confirms that depending on external force,
$\mathcal{K}$ as defined below equation (\ref{eq:astro_F_2_final}) may in principle destabilize
plane shear flows. 
Now from \S 2.1, for a 10 solar mass central object, $\mathcal{K}=0.1$, if the floating grains' 
number density is of the order of $5\times 10^{-7}$ cm$^{-3}$, which is a very small fraction compared to total 
number density of a protoplanetary accretion disc, hence quite viable. In reality, $Re$ for an accretion disc is 
several orders of magnitude higher than 1000, hence the required $\mathcal{K}$ for instability 
could be much smaller (see below for a more concrete description).
Nevertheless, in rest of the paper, we concentrate on equations (\ref{eq:orr_sommerfeld_eq}) and
(\ref{eq:squire_eq}) and their recasting forms without assuming any form of $\Gamma_1$ and $\Gamma_2$.

\subsection{Linear analysis}
\label{sec:lin-ana}

In the linear regime, equation (\ref{eq:mod_landau_eq}) becomes 
\begin{eqnarray}
\frac{dA_t}{dt} = \sigma_i A_t -\mathcal{N}.
\label{eq:mod_landau_eq_lin}
\end{eqnarray}
From equations (\ref{eq:N}), (\ref{eq:exp_for_Gamma'}) (and also from equation (\ref{eq:Gamma'_large_t})), it is not 
difficult to understand that at large $t$, $\mathcal{N}$ becomes constant over time, which is depicted 
in Fig.~\ref{fig:N_vs_t_R_500_1000_2000}. Now, the solution for equation (\ref{eq:mod_landau_eq_lin}) is 
\begin{eqnarray}
A_t = -\frac{1}{\mathcal{D}_t - \sigma_i}\mathcal{N} + Ce^{\sigma_i t}, 
\end{eqnarray}
where $C$ is an integration constant and $|A_t|$ becomes $|\mathcal{N}|/|\sigma_i|$ at large  $t$, i.e. 
when $||\mathcal{D}_t||/|\sigma_i|<1,$ and $\sigma_i$ is negative. The important point here is that the saturation of 
$|A_t|$ does not 
depend on the initial value of $|A_t|$. From wherever we start, $|A_t|$ reaches $|\mathcal{N}|/|\sigma_i|$ (see below for details).


\begin{figure}
\includegraphics[width=\columnwidth]{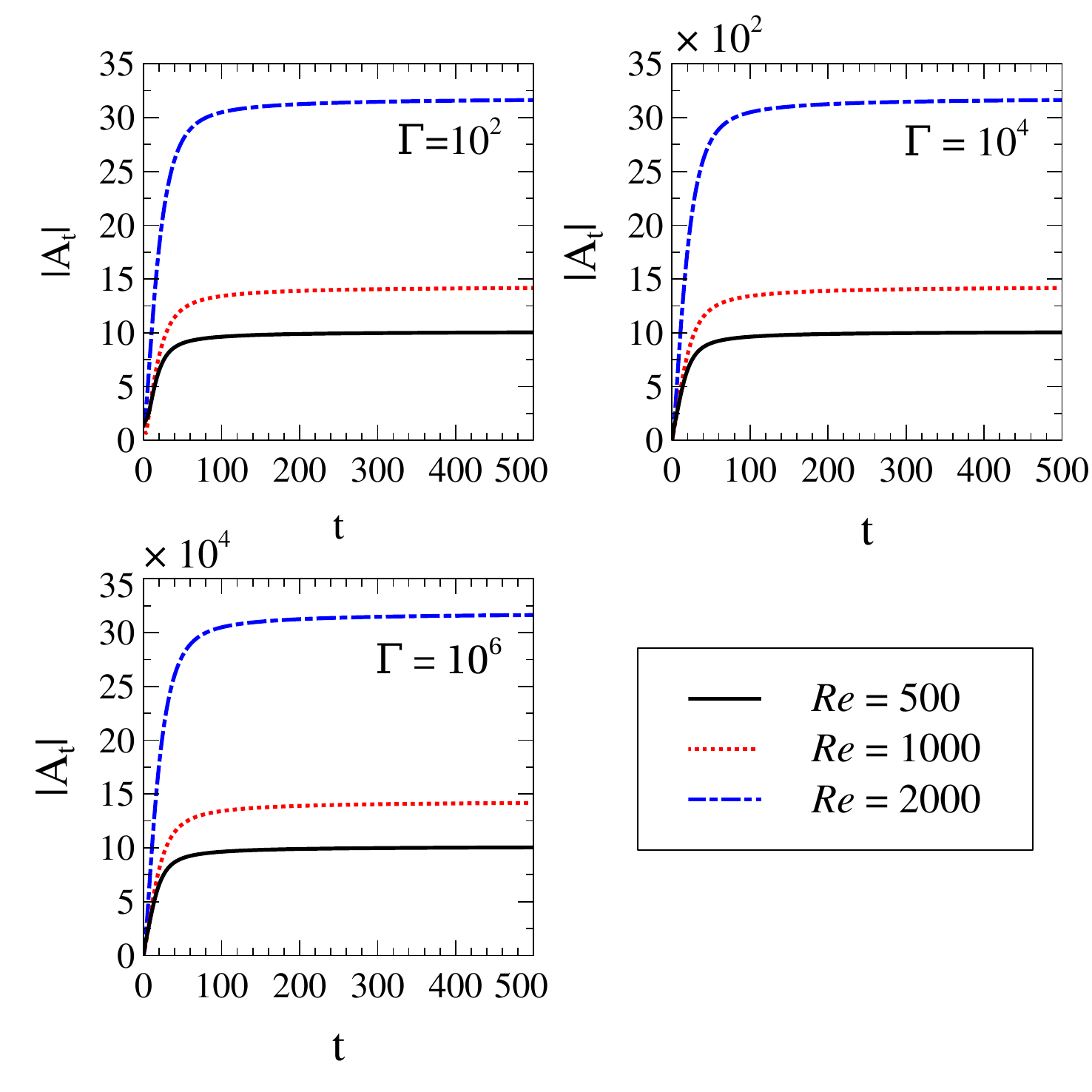}
\caption{Variation of $|A_t|$ as a function of $t$ for three sets of $Re$ and $\Gamma$ with $k_y = k_z=1$ for linear 
analysis in 
the Keplerian flow ($q=1.5$).}
\label{fig:A_t_vs_t_lin_3_diff_R_3_diff_gamma}
\end{figure}

\begin{figure}
\includegraphics[width=\columnwidth]{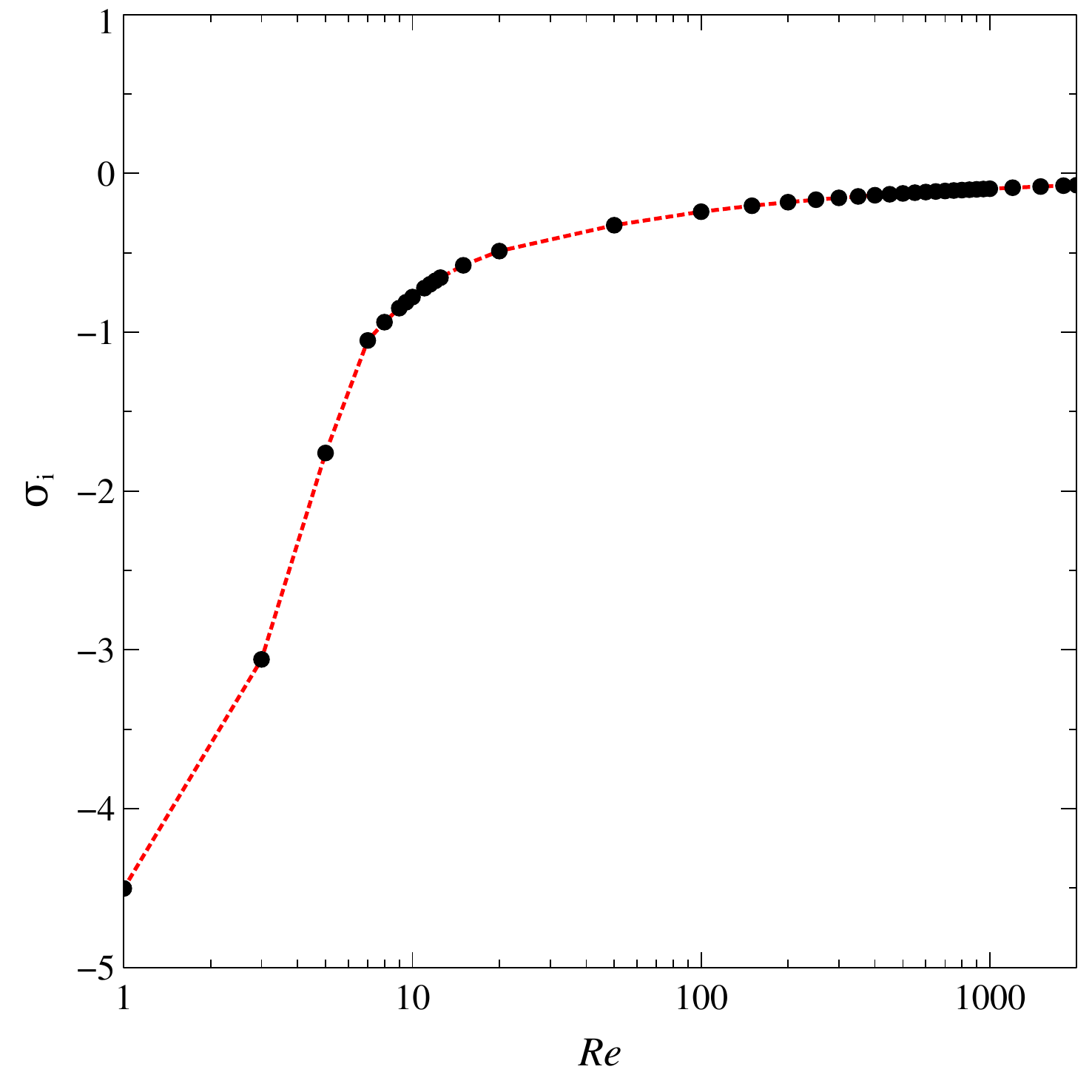}
\caption{Variation of $\sigma_i$ as a function of $Re$ for $k_y = k_z=1$ in the Keplerian flow.}
\label{fig:R_vs_sigma_i_kep}
\end{figure}

Fig.~\ref{fig:A_t_vs_t_lin_3_diff_R_3_diff_gamma} shows the variation of $|A_t|$ as a function of $t$ for various values 
of $Re$ and 
$\Gamma$. From equation (\ref{eq:Gamma}) we can fix $\Gamma$ by fixing the position, i.e. $x,y$ and $z$, and choosing 
$\Gamma_1$ and 
$\Gamma_2$. Fig.~\ref{fig:A_t_vs_t_lin_3_diff_R_3_diff_gamma} also suggests the scaling relation between saturated 
$|A_t|$ and 
$\Gamma$ to be
\begin{eqnarray}
 |A_t|\propto \Gamma
 \label{eq:A_t_Gamma}
\end{eqnarray}
for a fixed $Re$.

Now from Fig.~\ref{fig:R_vs_sigma_i_kep}, we see that $|\sigma_i|$ becomes smaller and smaller as $Re$ increases. On the 
other hand, 
Fig.~\ref{fig:N_vs_t_R_500_1000_2000} shows that the saturated value of $|\mathcal{N}|$ becomes larger for larger $Re$. 
Therefore the 
saturated value of $|A_t|$, i.e. $|\mathcal{N}|/|\sigma_i|$, becomes larger for larger $Re$. Therefore for $Re = 
10^{10}$, the 
saturated value of $|A_t|$ will be huge and this in fact leads the perturbations to be highly nonlinear, which further 
could make the 
flow turbulent. The emergence of nonlinearity and hence further the turbulence, in this context, can be interpreted in 
the following way also. At 
the 
linear regime, the amplitude $|A_t|$ is so small that $|\sigma_i A_t| \gg |p|A_t|^2A_t|$. If $A_t$ evolves in such a way 
that linear and 
nonlinear terms become equivalent, i.e. $|A_t|^2  \sim |\sigma_i|/|p|$, then the nonlinear part comes into the picture. 
Now if $Re$ 
increases, $|\sigma_i|$ decreases and $|p|$ increases. Thus, $|A_t|$ for the onset of nonlinearity decreases as $Re$ 
increases. For $Re=500$, $|\sigma_i|\sim 10^{-1}$ and $|p| \sim 10^{-4}$ for the Keplerian flow. This leads to $|A_t|\sim 
33$ for the onset 
of 
nonlinearity in the system. From Fig.~\ref{fig:A_t_vs_t_lin_3_diff_R_3_diff_gamma}, we notice that 
for $Re=500$ at $\Gamma = 10^2$, the saturation value 
of $|A_t|$ is about 8, while at $\Gamma = 10^4$ the saturated $|A_t|$ is about 800.  On the other hand, for $Re = 2000$, 
$|\sigma_i|\sim 10^{-2}$ and $|p| \sim 10^{-3}$, and the flow starts to become nonlinear at around $|A_t| \sim 3.33$ for 
the Keplerian 
flow. 
Fig.~\ref {fig:A_t_vs_t_lin_3_diff_R_3_diff_gamma} suggests that even $\Gamma = 10^2$ could bring nonlinearity into the 
system for 
$Re=2000,$ as the 
saturation of $|A_t|$ therein occurs at around $|A_t| \sim 32$ which is almost 10 times the required value of 
$|A_t|$ for onsetting nonlinearity in the system. Hence, with increasing $Re$,
required $\Gamma$ to lead to nonlinearity and plausible turbulence becomes
smaller and smaller. As $Re$ in accretion discs is quite huge ($\gtrsim 10^{14}$,
see, e.g., \citealt{2013PhLB..721..151M}), required $\Gamma$ is very tiny.
 
Nevertheless, the occurrence of nonlinearity in this regard is quite amazing. The saturation of $|A_t|$ has nothing to do 
with the 
initial amplitude of the perturbation. Hence any small disturbance could make the flow nonlinear at a time, having a 
lower bound: 
$t>1/|\sigma_i|$ (from the assumption $||\mathcal{D}_t||/|\sigma_i|<1$).

\begin{figure}
\includegraphics[width=\columnwidth]{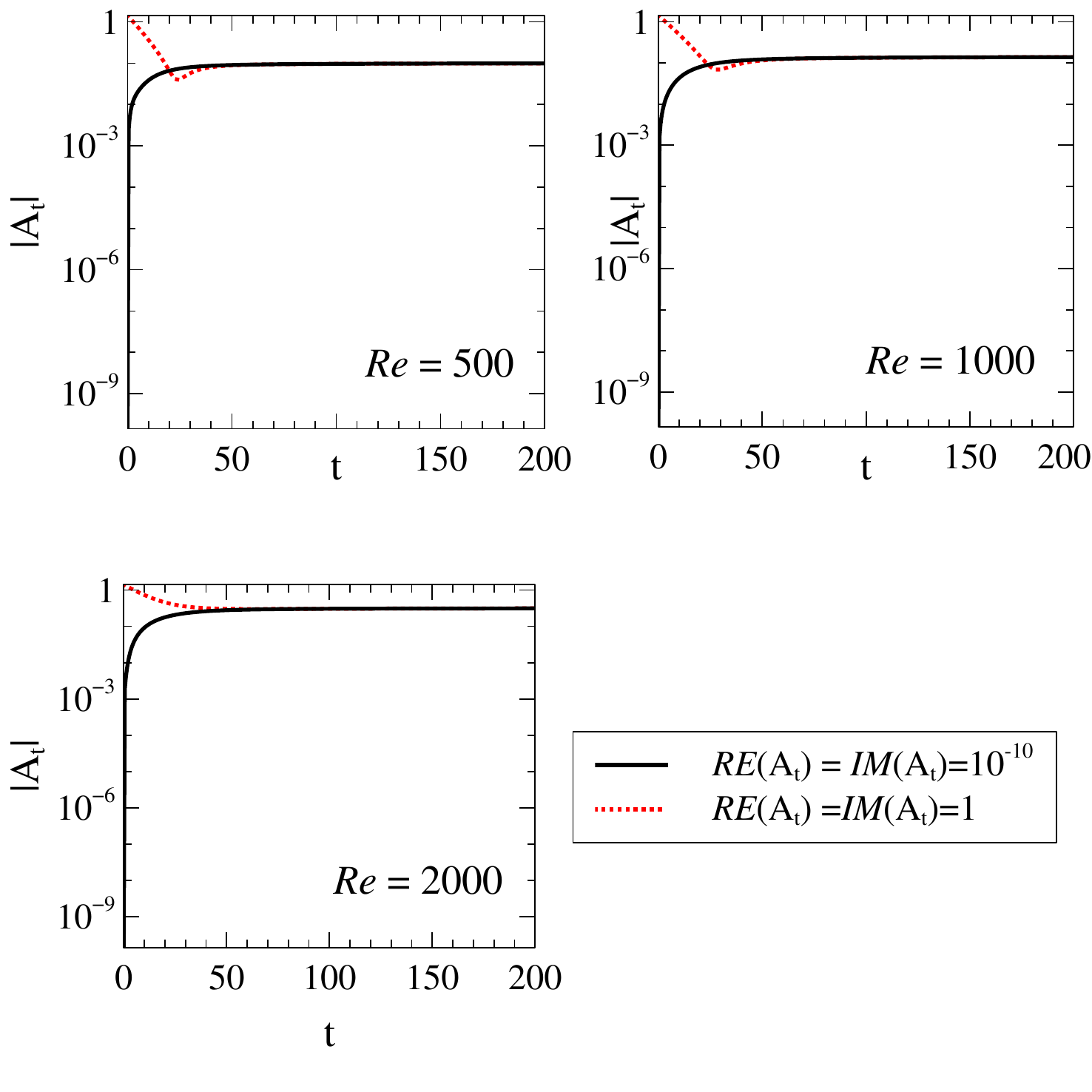}
\caption{Variation of $|A_t|$ as a function of $t$ for $\Gamma = 1$ at two different initial conditions 
with $k_y = k_z=1$ for three different values of $Re$. The two different initial conditions are real part ($RE(A_t)$) and 
imaginary part 
($IM(A_t)$) of 
$A_t$ to be 1 and 
$10^{-10}$.} 
\label{fig:A_t_vs_t_lin_gamma_1_diff_A_0}
\end{figure}

Fig.~\ref{fig:A_t_vs_t_lin_gamma_1_diff_A_0} shows the variation of $|A_t|$ as a function of $t$ for two initial 
conditions and for a 
particular $Re$, showing the same saturated value of $|A_t|$. For all three cases $\Gamma$ is 1. This confirms the 
independence of initial 
condition.

\begin{figure}
\includegraphics[width=\columnwidth]{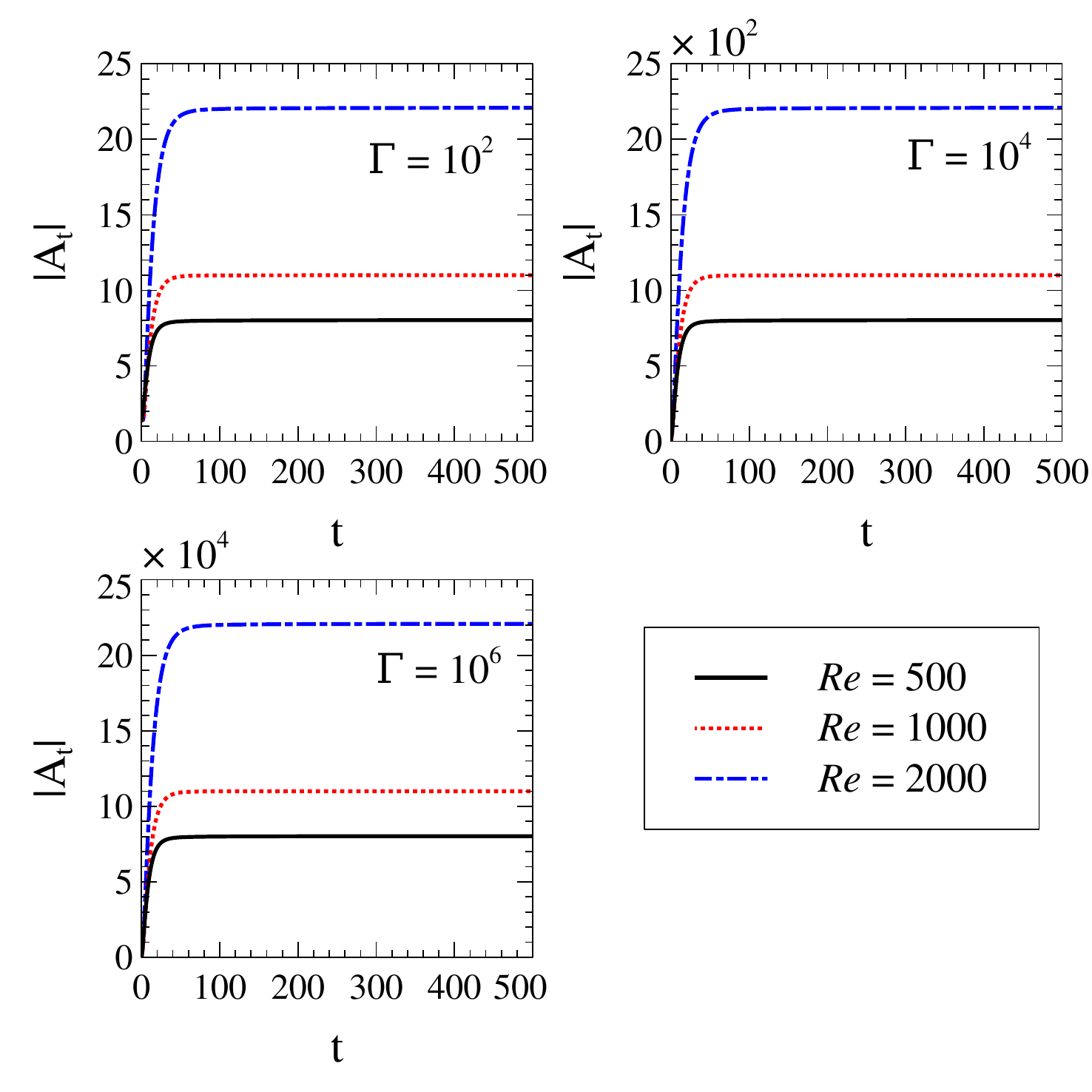}
\caption{Variation of $|A_t|$ as a function of $t$ for three sets of $Re$ and $\Gamma$ with $k_y = k_z=1$ for the linear 
analysis in plane 
Couette flow ($q \rightarrow \infty$).}
\label{fig:A_t_vs_t_palne_cou_lin_3_diff_R_3_diff_gamma}
\end{figure}

\begin{figure}
\includegraphics[width=\columnwidth]{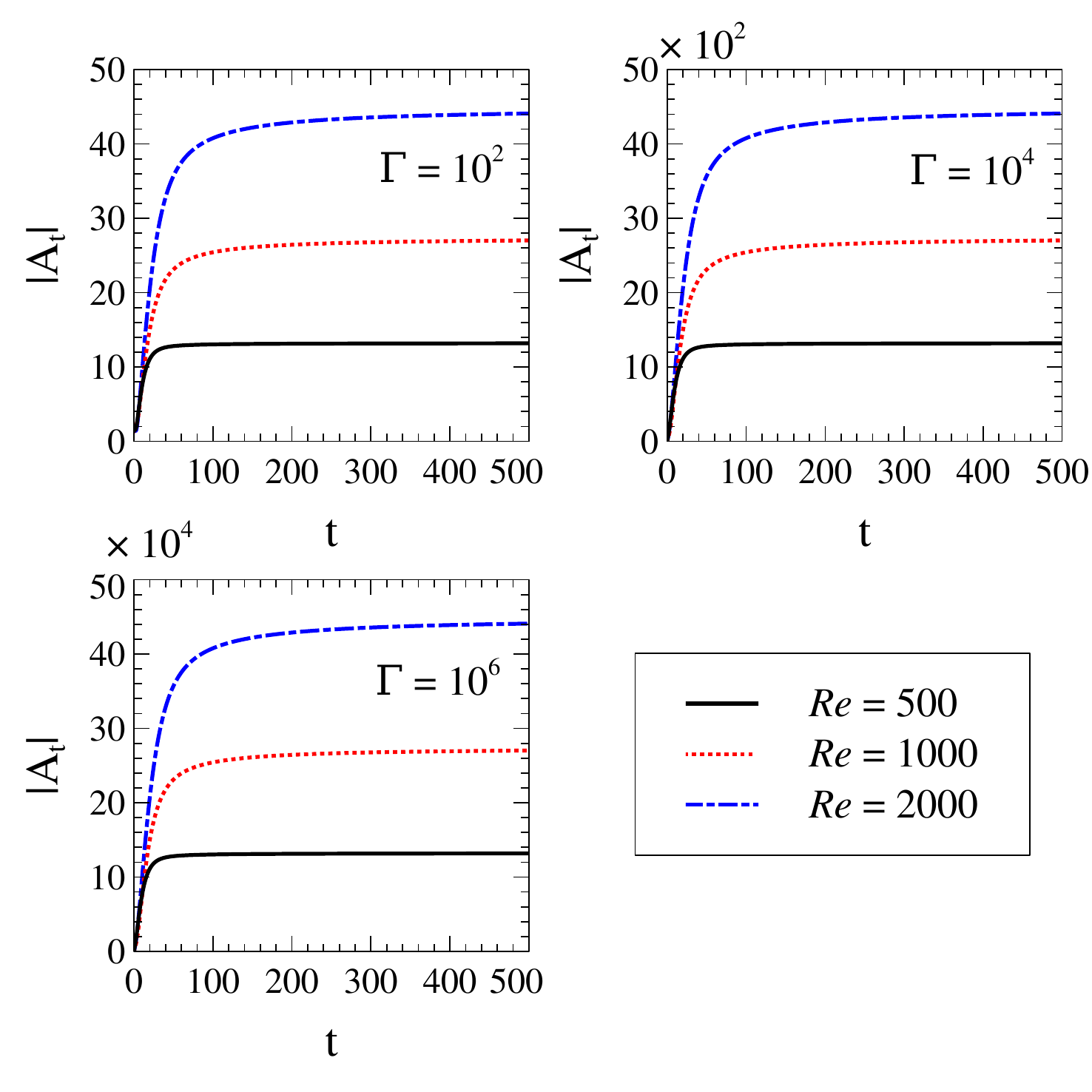}
\caption{Variation of $|A_t|$ as a function of $t$ for three sets of $Re$ and $\Gamma$ with $k_y = k_z=1$ for the linear 
analysis in
constant angular momentum flow ($q=2$).}
\label{fig:A_t_vs_t_q_2_lin_3_diff_R_3_diff_gamma}
\end{figure}

Figs.~\ref{fig:A_t_vs_t_palne_cou_lin_3_diff_R_3_diff_gamma} and \ref{fig:A_t_vs_t_q_2_lin_3_diff_R_3_diff_gamma} show 
the 
variation of $|A_t|$ as a function of $t$ at various $Re$ and $\Gamma$ for plane Couette and constant angular momentum 
flows. All the 
results are similar to those of the Keplerian flow. The equation (\ref{eq:A_t_Gamma}) holds for both the cases. But 
interestingly, the 
saturated value of $|A_t|$ for a particular $Re$ and $\Gamma$ is the largest for constant angular momentum flow and the 
smallest 
for plane Couette flow among the three kinds of flow. Although the eigenspectra for constant angular momentum flow and 
plane Couette flow 
are same, the eigenmodes corresponding to the same eigenvalues for these two 
flows are not the same as nonzero matrix elements in $\mathcal{L}$ in equation (\ref{eq:the_L_matrix}) are not the same 
for both the flows. 
Equation (\ref{eq:N}) shows the dependence of $\mathcal{N}$ on the adjoint eigenmodes of $\mathcal{L}$. This is the 
reason behind obtaining 
different evolution of $|A_t|$ for the constant angular momentum flow and plane Couette flow. Now we 
interpret Figs.~\ref{fig:A_t_vs_t_lin_3_diff_R_3_diff_gamma} and \ref{fig:A_t_vs_t_q_2_lin_3_diff_R_3_diff_gamma} in 
terms of epicyclic 
frequency which is given by
\begin{equation}
\kappa = \sqrt{2(2-q)}\Omega,
\label{eq:epicyclic_freq}
\end{equation}
where $\Omega$ is the angular frequency of the fluid parcel. The real value of $\kappa$ indicates the oscillation about 
the mean position 
of 
the fluid parcel, while the imaginary value of $\kappa$ indicates unstable fluid parcel after it is perturbed. However, 
$\kappa$ is zero 
for $q=2$ (i.e. constant angular momentum flow) and some positive real number for $q=1.5$ (i.e. the Keplerian flow). 
Hence, constant angular 
momentum flow is a marginally stable flow and the Keplerian flow is a well stable flow. From 
Figs.~\ref{fig:A_t_vs_t_lin_3_diff_R_3_diff_gamma} 
and \ref{fig:A_t_vs_t_q_2_lin_3_diff_R_3_diff_gamma}, we notice that the saturated value of $|A_t|$ for constant angular 
momentum flow is 
larger than that for the Keplerian flow. The order of nonlinearity is, therefore, higher in the constant angular momentum 
flow than that in 
the Keplerian flow and, thence, plausibility of turbulence. 

\begin{figure}
\includegraphics[width=\columnwidth]{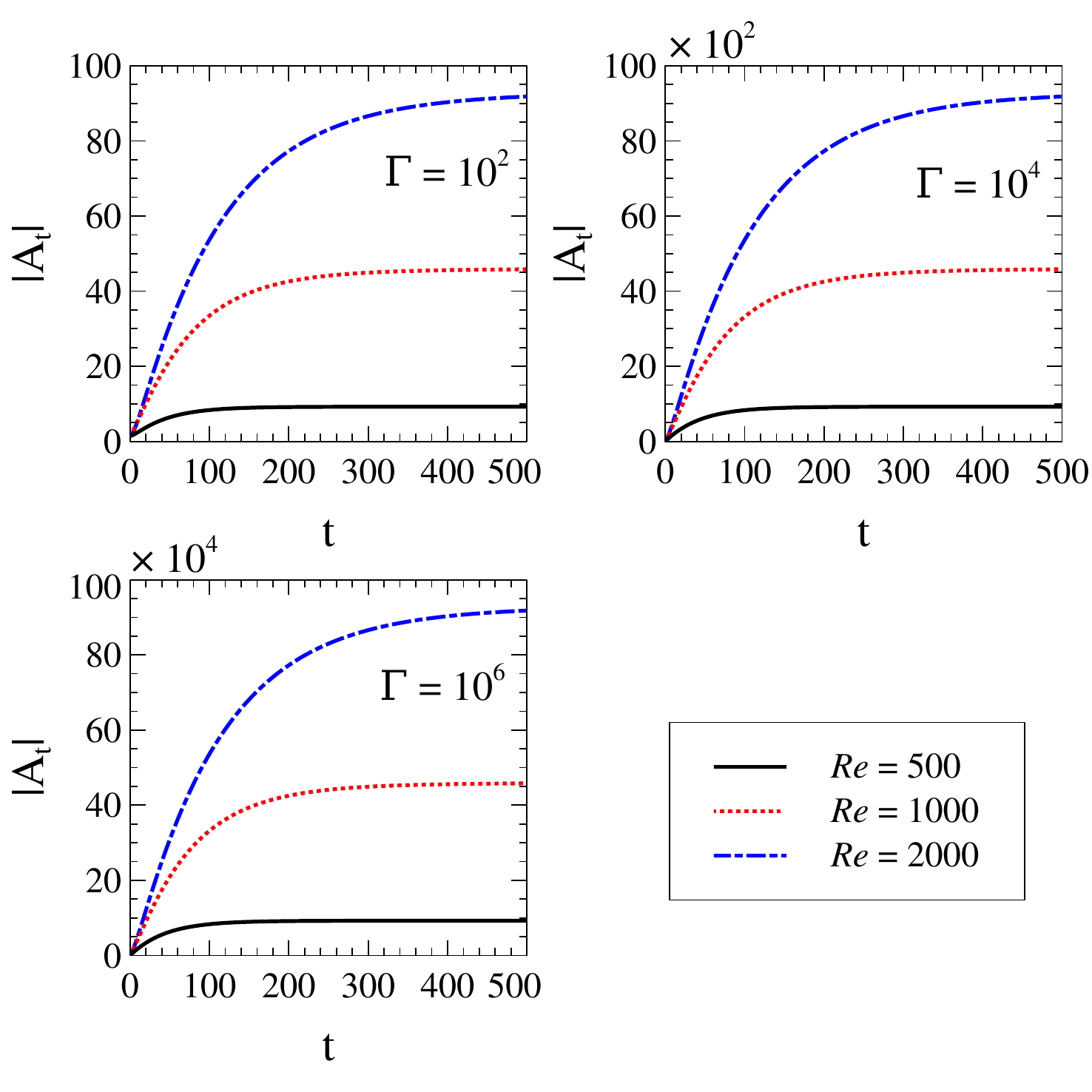}
\caption{Variation of $|A_t|$ as a function of $t$ for three sets of $Re$ and $\Gamma$ with $k_y = 0.1$ and $k_z=1$ for 
linear analysis in 
the Keplerian flow.}
\label{fig:A_t_vs_t_lin_3_diff_R_3_diff_gamma_ky_10_-1_kz_1}
\end{figure}

\begin{figure}
\includegraphics[width=\columnwidth]{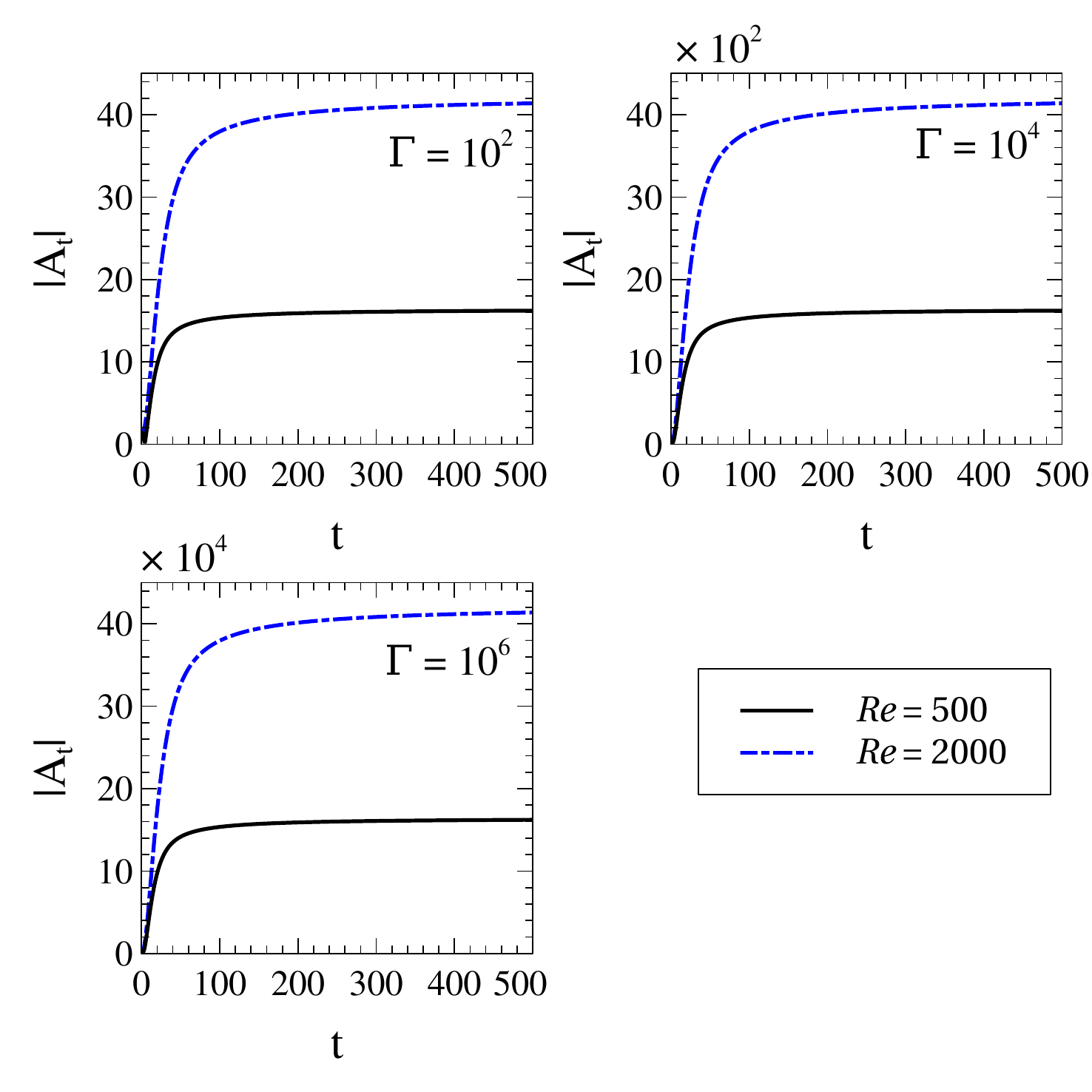}
\caption{Variation of $|A_t|$ as a function of $t$ for various $Re$ and $\Gamma$ with $k_y = 1$ and $k_z=0.1$ for linear 
analysis in the 
Keplerian flow.}
\label{fig:A_t_vs_t_lin_3_diff_R_3_diff_gamma_ky_1_kz_10_-1}
\end{figure}

\begin{figure}
\includegraphics[width=\columnwidth]{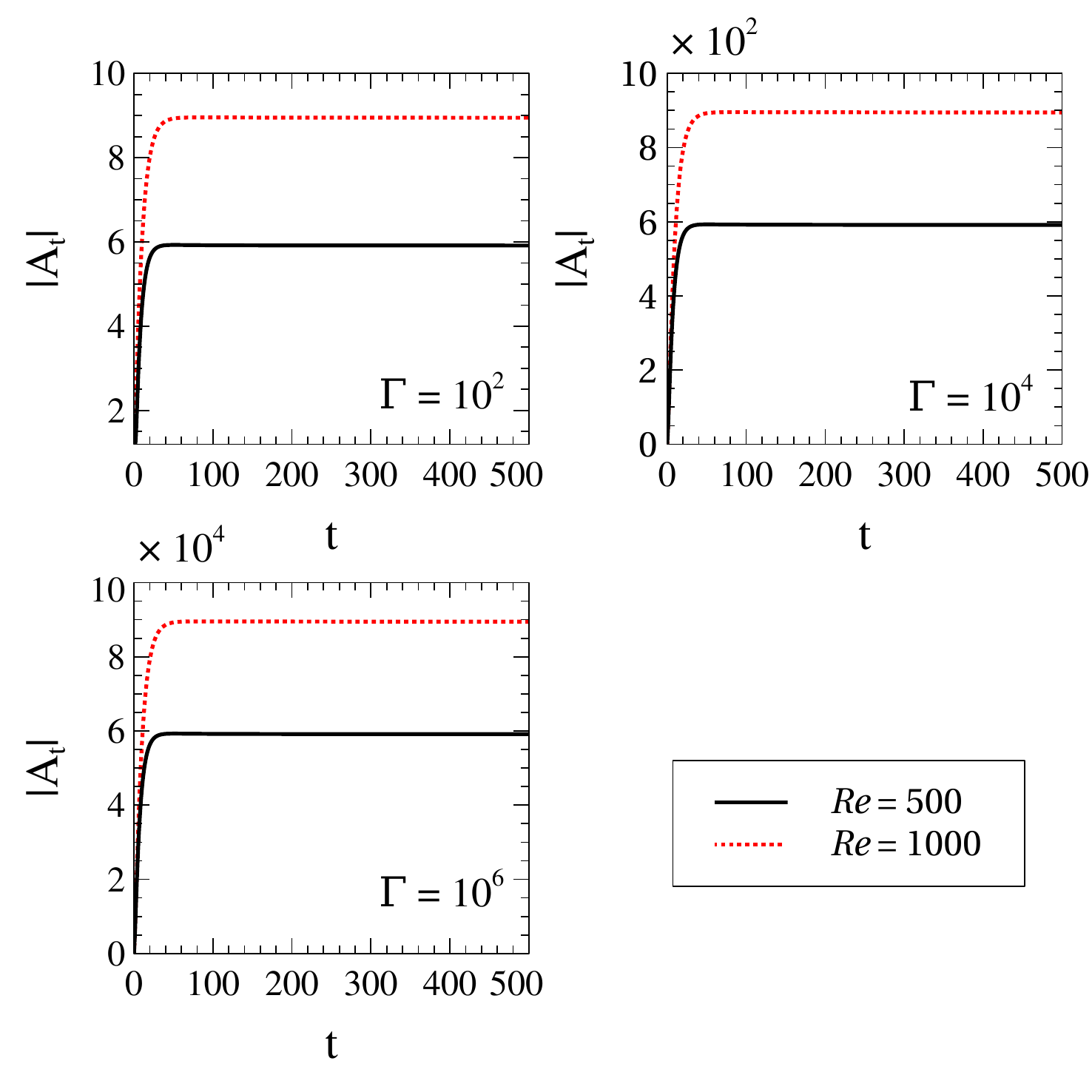}
\caption{Variation of $|A_t|$ as a function of $t$ for various $Re$ and $\Gamma$ with $k_y = 1$ and $k_z=0$ for linear 
analysis.}
\label{fig:A_t_vs_t_lin_3_diff_R_3_diff_gamma_ky_1_kz_0}
\end{figure}

Fig.~\ref{fig:A_t_vs_t_lin_3_diff_R_3_diff_gamma_ky_10_-1_kz_1} shows the variation of $|A_t|$ as a function of $t$ at 
various $Re$ and 
$\Gamma$ but for $k_y = 0.1$ and $k_z = 1$ in the Keplerian flow. This case is a representative example exhibiting 
vertically dominated 
perturbation. It also shows that the saturated $|A_t|$ is larger compared to that of the $k_y = k_z = 1$ case, 
when the time to saturate also turns out to be longer. This is due to the fact that $|\sigma_i|$ is smaller for this case 
than that for 
$k_y = k_z = 1$ case for a fixed $Re$. 
Similarly, if we make the perturbation more planer, i.e. decrease $k_z$ for a fixed $Re$ and $q$, the saturated value of 
$|A_t|$ increases, compared to the $k_y=k_z=1$ case, but the time to saturate also turns out to be shorter. Fig.~\ref{fig:A_t_vs_t_lin_3_diff_R_3_diff_gamma_ky_1_kz_10_-1} 
depicts this phenomena for the Keplerian flow with $k_y = 1$ and 
$k_z = 0.1$. If we make $k_z = 0$, the perturbations are entirely two-dimensional and the rotational effect is completely 
suppressed. The 
variation of $|A_t|$, therefore, will no longer depend on $q$. 
Fig.~\ref{fig:A_t_vs_t_lin_3_diff_R_3_diff_gamma_ky_1_kz_0} shows the 
variation of $|A_t|$ as a function of $t$ for various $Re$ and $\Gamma$ for two-dimensional perturbation, i.e. $k_y = 1$ 
and $k_z = 0$, when the time to saturate is shortest. Note importantly that for each $q$, there is an optimum set of
$k_y$ and $k_z$, giving rise to the best least stable mode and growth, whose imaginary part of eigenvalue
decreases with decreasing $q$ below $2$. However,
at present, we do not concentrate on the optimum set(s) of $k_y$ and $k_z$. Hence,
stabilizing effect with respect to rotation is not reflected here. 
Nevertheless, it is evident that as the perturbation varies from
vertical to planner, the time to saturate the growth becomes shorter.

\subsection{Nonlinear analysis}
If there is no extra force involved in the system, then equation (\ref{eq:mod_landau_eq}) becomes the usual Landau 
equation, which is
\begin{eqnarray}
 \frac{dA_t}{dt} = \sigma_i A_t + p|A_t|^2A_t,
 \label{eq:original_landau_eq}
\end{eqnarray}
which can be further recast to
 \begin{eqnarray}
\frac{d|A|^2}{dt}  = k_1 |A|^2 + k_2|A|^4,
\label{eq:landau_eq}
\end{eqnarray}
where $A$ is the amplitude of the nonlinear perturbations for the corresponding system, $k_1$ is 2$\sigma_i$ and 
$k_2$ is the real part of $2p$, i.e. 2$p_r$. Its solution is 
\begin{eqnarray}
 |A|^2 = \frac{A_0^2}{-\frac{k_2}{k_1}A_0^2+\left(1+\frac{k_2}{k_1}A_0^2\right)e^{-k_1t}}.
 \label{eq:sol_landau_eq}
\end{eqnarray}
If both $k_1$ and $k_2$ are positive, then we can find a particular time (by making the denominator of equation 
(\ref{eq:sol_landau_eq}) 
to 0),
\begin{eqnarray}
 t = -\frac{1}{k_1}ln\left(\frac{k_2A_0^2}{k_1+k_2A_0^2}\right)
 \label{eq:t}
\end{eqnarray}
at which $|A|$ diverges. Therefore, in this case, the system becomes highly nonlinear and we have to consider all kinds 
of 
nonlinear effects. Thus the system is expected to become turbulent rapidly.

\begin{figure}
\includegraphics[width=\columnwidth]{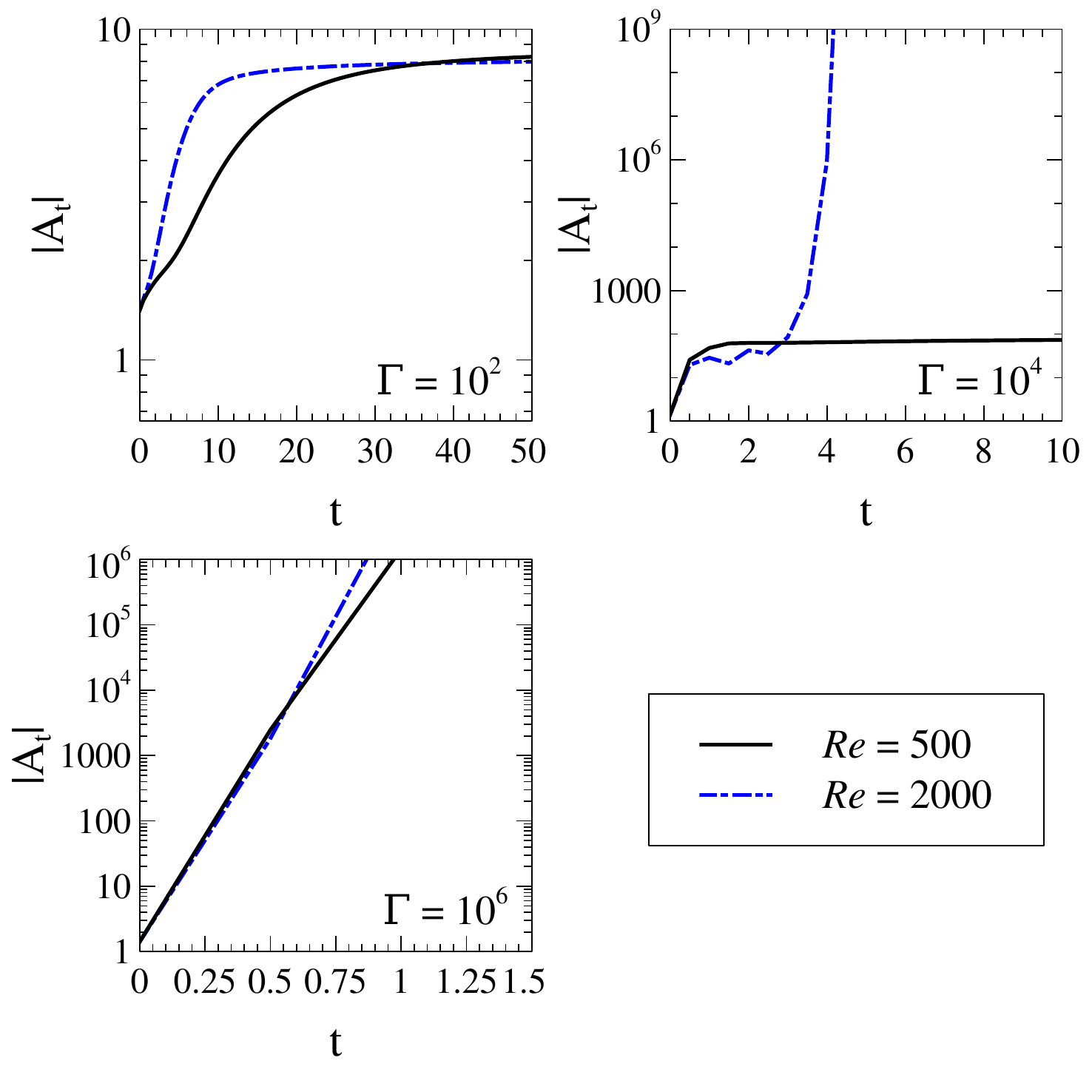}
\caption{Variation of $|A_t|$ as a function of $t$ for different $Re$ and $\Gamma$ with $k_y = k_z=1$ for nonlinear 
analysis in the 
Keplerian flow.}
\label{fig:A_t_vs_t_nonlin_3_diff_R_3_diff_gamma}
\end{figure}

However, the presence of extra force makes it very difficult for us to have a compact analytical solution like equation 
(\ref{eq:sol_landau_eq}). 
Therefore, we venture for numerical solutions of equation (\ref{eq:mod_landau_eq}) for different parameters such as 
$Re$ and $\Gamma$. 
Fig.~\ref{fig:A_t_vs_t_nonlin_3_diff_R_3_diff_gamma} shows the solution of equation (\ref{eq:mod_landau_eq}), describing 
the variation of 
$|A_t|$ from equation (\ref{eq:mod_landau_eq}) as a function of $t$ for  $Re = $ 500 and 2000 for different 
$\Gamma$ in the Keplerian 
flow. 
We notice that $\Gamma$ plays an important role. $|A_t|$ saturates for $\Gamma=10^2$ beyond a certain time. However, as 
$\Gamma$ increases 
to
$10^4$, we see that $|A_t|$ diverges for $Re=2000$ at a certain time, but not for $Re = 500$. As the strength of the 
external force, i.e. 
$\Gamma$, further increases to $10^6$, we see that $|A_t|$ diverges at a smaller time and even at a smaller $Re$. 
    
\begin{figure}
\includegraphics[width=\columnwidth]{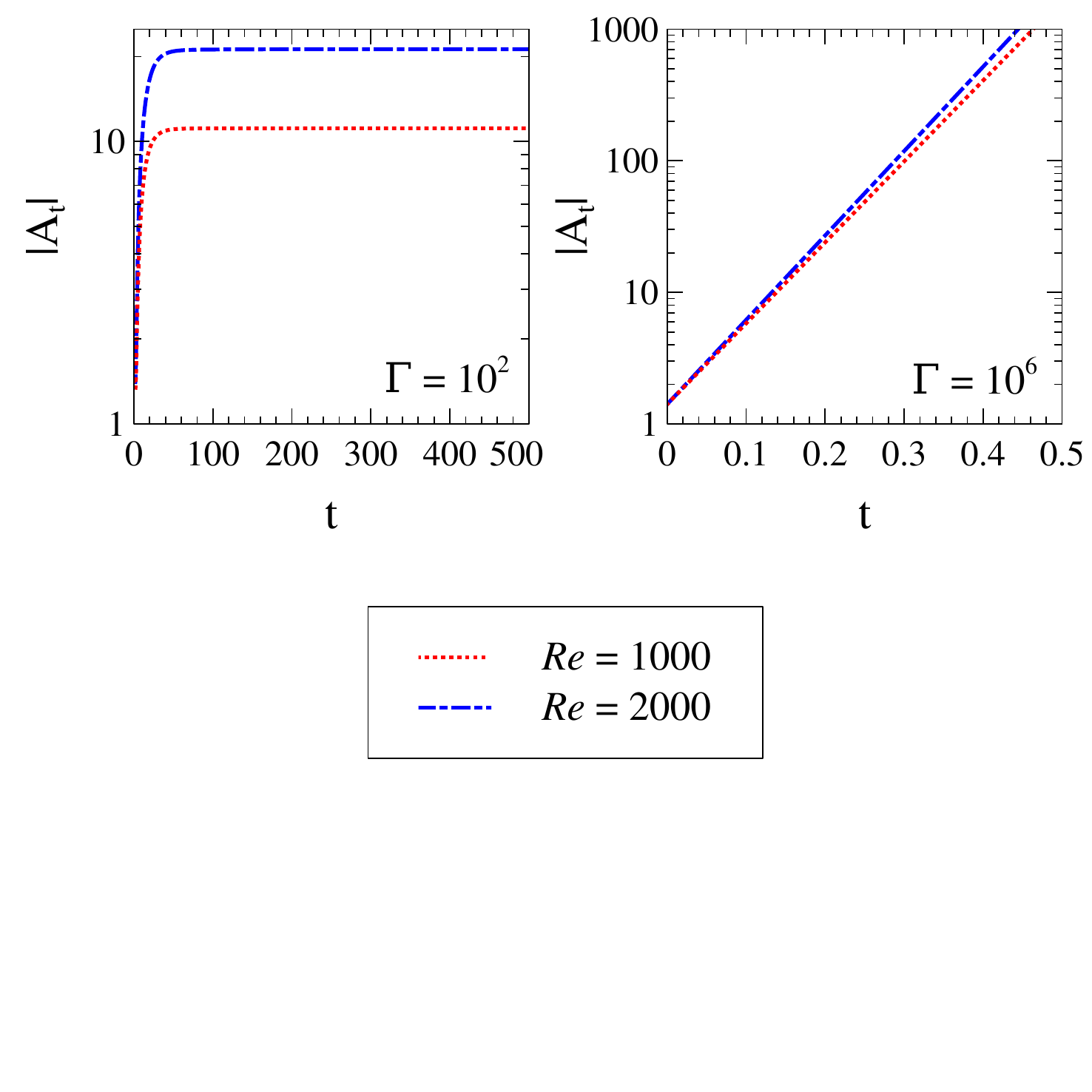}
\caption{Variation of $|A_t|$ as a function of $t$ for different of $Re$ and $\Gamma$ with $k_y = k_z=1$ for nonlinear 
analysis in 
plane Couette flow.}
\label{fig:A_t_vs_t_plane_cou_nonlin_3_diff_R_3_diff_gamma}
\end{figure}

Fig.~\ref{fig:A_t_vs_t_plane_cou_nonlin_3_diff_R_3_diff_gamma} shows the variation of $|A_t|$ as a function of $t$ for 
different $Re$ for 
plane Couette flow. The results are quite similar to those for the Keplerian flow.

\begin{figure}
\includegraphics[width=\columnwidth]{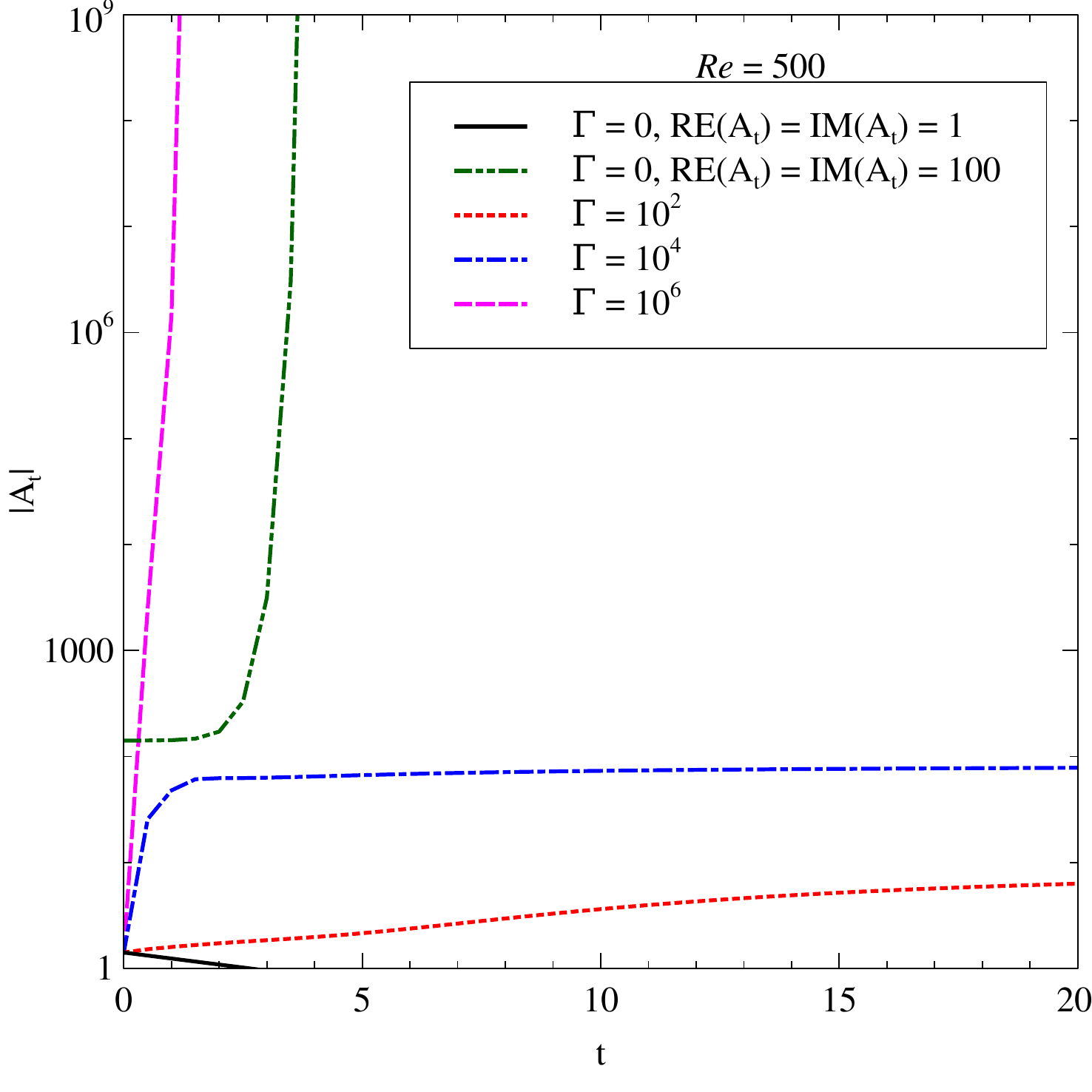}
\caption{Variation of $|A_t|$ as a function of $t$ with $k_y = k_z=1$ for nonlinear analysis in the Keplerian flow for 
$Re=500$ and four 
different $\Gamma$. For $\Gamma\neq 0$, initial condition is RE($A_t$) = IM($A_t$) = 1.}
\label{fig:A_t_vs_t_nonlin_landau_modi_landau_diff_gamma_R_500}
\end{figure}

Nevertheless, in our case, $k_1$ in equation (\ref{eq:landau_eq}) is negative. It makes the problem more interesting if 
$k_2>0$. 
In the absence of force, if the initial amplitude of 
the perturbation $A_0$ is larger than the threshold amplitude,
\begin{eqnarray}
|A| = A_i = \sqrt{\frac{-k_1}{k_2}},
\label{eq:threshold_A}
\end{eqnarray}
then it is well-known that (see, e.g., \citealt{1970JFM....40...97E,2004hyst.book.....D} for plane Couette flow,
\citealt{2011MNRAS.414..691R} for the Keplerian discs) there will be a time $t$, as given in equation (\ref{eq:t})
(with suitable sign of $k_1$ and $k_2$ in mind), at which the solution 
diverges. This is shown in the 
Fig.~\ref{fig:A_t_vs_t_nonlin_landau_modi_landau_diff_gamma_R_500} with a dashed-short-dashed (green) growing line 
starting from finite $|A_t|$ for $Re=500$ and $k_y = k_z=1$, 
whereas the solid (black) fast decaying line indicates the result with smaller $A_0$. Other three 
curves, starting from the same smaller $|A_t|=|A_0|$, are showing the variation of $|A_t|$ as a function of $t$ in the very presence of the extra force. The pattern of 
turbulence 
during its onset in the absence of extra force but with a finite initial amplitude of perturbation at $325\lesssim Re\lesssim 380$ for plane Couette flow was simulated by 
\cite{Duguet_2010}. 
In our case shown in Fig.~\ref{fig:A_t_vs_t_nonlin_landau_modi_landau_diff_gamma_R_500} by the dashed-short-dashed (green) line, we also 
see the diverging nature of amplitude of the nonlinear perturbation beyond a certain time in the absence of extra 
force, but in the presence of Coriolis force (which is a stabilizing effect), only with finite 
initial amplitude of perturbation. This implies the turbulent nature of the 
flow. It is apparent that the onset of the nonlinearity depends on the initial amplitude 
of perturbation in the absence of the force, but it does not depend on the same in the presence of force. The divergence 
of $|A_t|$ and 
hence the onset of nonlinearity and plausible turbulence depends only on the strength of the force,
as shown by dashed (magenta) line, compared to dot-dashed (blue) and dotted (red) lines, in Fig.~\ref{fig:A_t_vs_t_nonlin_landau_modi_landau_diff_gamma_R_500}. 
The presence of 
$\Gamma$ with negative $k_1$ ($\sigma_i$) 
is equivalent to the Landau equation and solution with $\Gamma = 0$ and $k_1$ and $k_2$ both positive. With a 
suitable strength of force, $|A_t|$ 
diverges quicker than that without force.

\subsubsection{Plane Couette flow and bounds on parameters}

\begin{figure}
\includegraphics[width=\columnwidth]{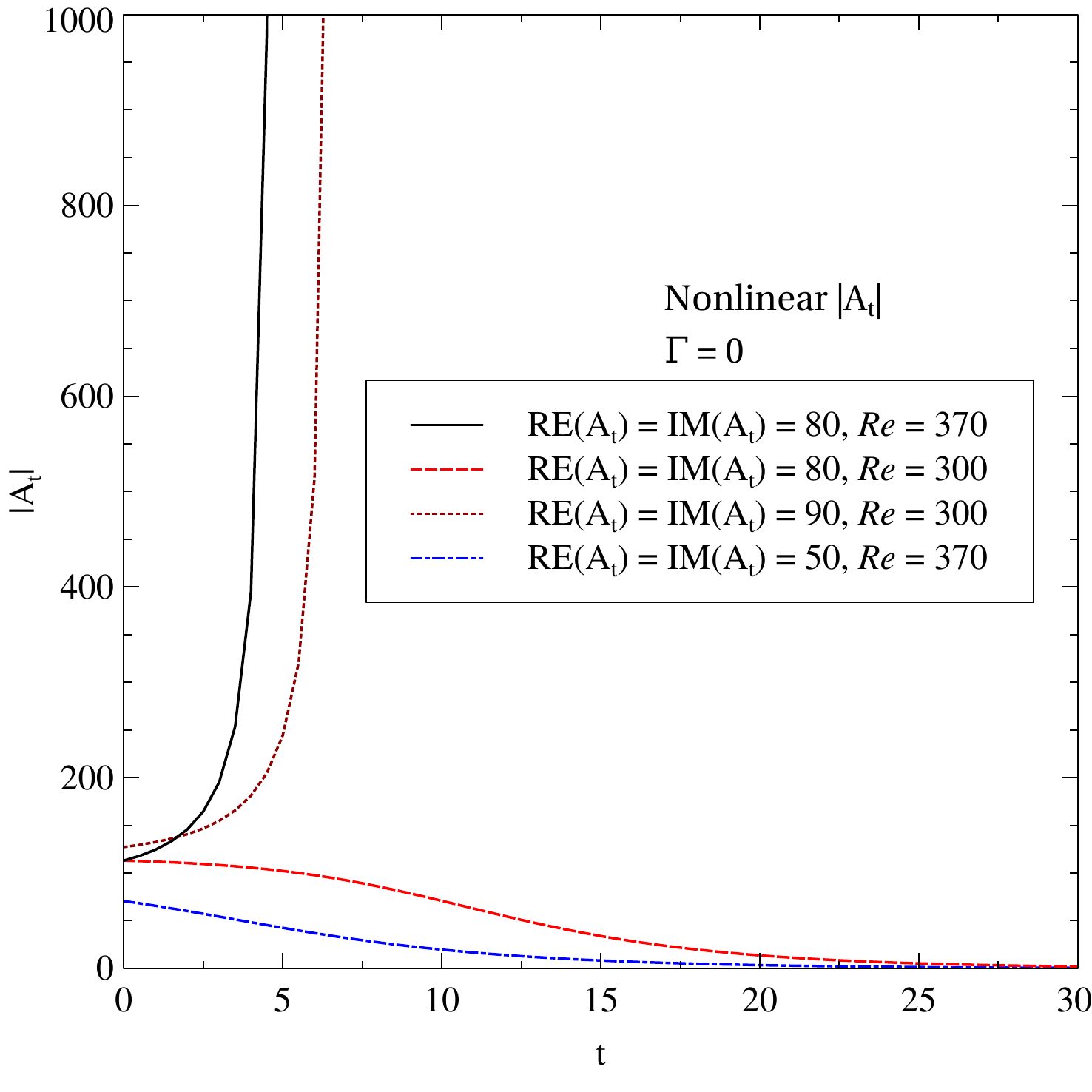}
\caption{Variation of $|A_t|$ as a function of $t$ with $k_y = k_z=1$ for nonlinear analysis in 
plane Couette flow without force for different $Re$ and initial conditions.}
\label{fig:A_t_vs_t_plane_cou_nonlin_R_300_370_no_force_diff_ini_cond}
\end{figure}

Similar results as above are obtained for plane Couette flow, in accordance with the simulation 
by \cite{Duguet_2010}. Fig. \ref{fig:A_t_vs_t_plane_cou_nonlin_R_300_370_no_force_diff_ini_cond}
shows that for a given initial amplitude of perturbation, while $|A_t|$ decays with time for $Re=300$, 
increasing $Re$ to 370 leads to diverging $|A_t|$ at a finite time. Also, for a given $Re$, a smaller 
initial amplitude of perturbation depending on $Re$, makes $|A_t|$ decaying with time. While a very large 
initial amplitude might make $|A_t|$ diverging even at $Re=300$, that situation might be naturally
implausible or equivalent to external forcing. That is perhaps the reason that \cite{Duguet_2010} 
found plane Couette flow laminar for $Re<324$. If $|A_t|$ should be finite for $Re<324$, initial
amplitude should have an upper bound, e.g. $\lesssim 80$, perhaps larger initial amplitude is naturally
implausible.

\begin{figure}
\includegraphics[width=\columnwidth]{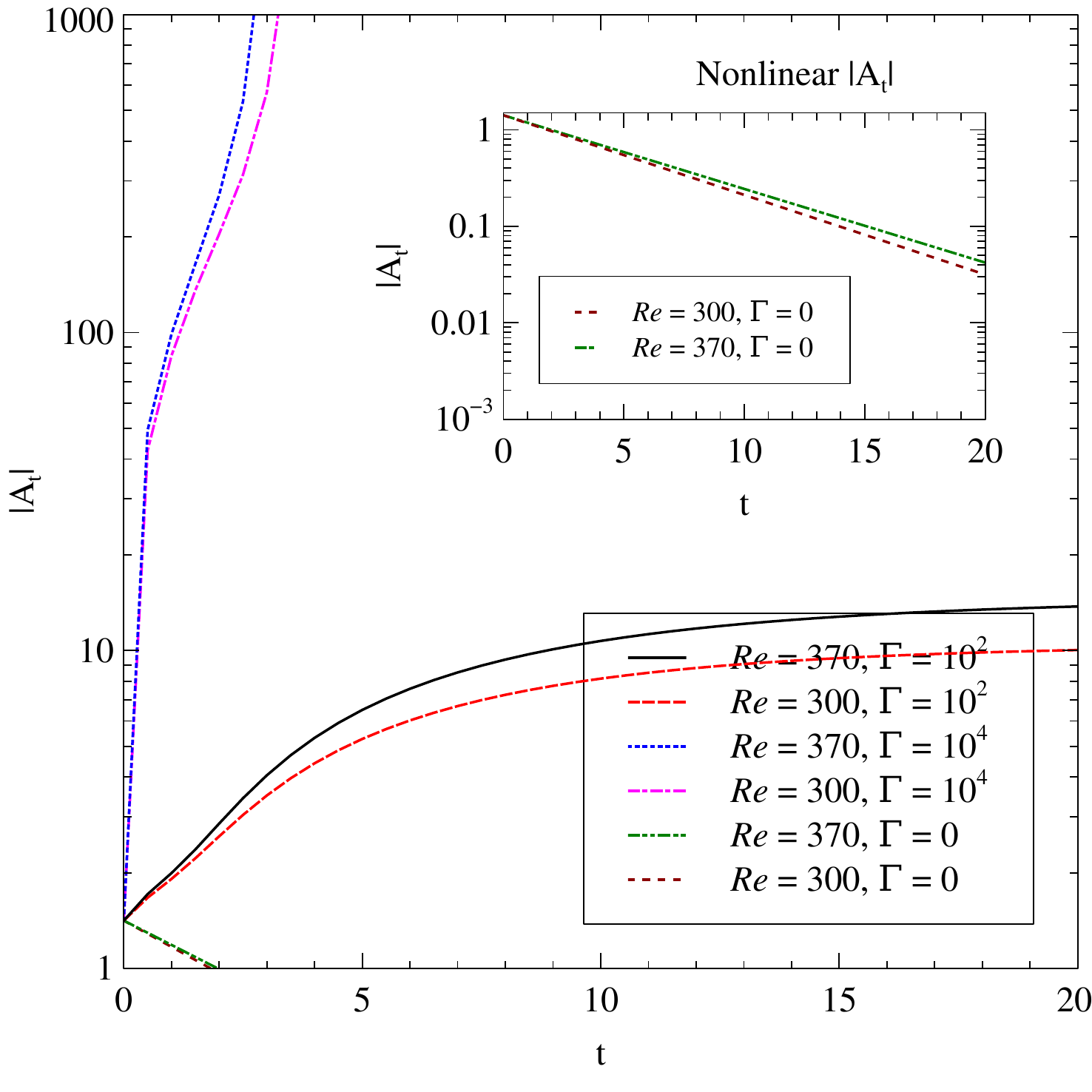}
\caption{Variation of $|A_t|$ as a function of $t$ with $k_y = k_z=1$ for nonlinear analysis in 
plane Couette flow for different forces and $Re$ at a fixed initial condition RE($A_t$) = IM($A_t$) = 1.}
\label{fig:A_t_vs_t_plane_cou_nonlin_R_300_370_force_no_force}
\end{figure}

\begin{figure}
\includegraphics[width=\columnwidth]{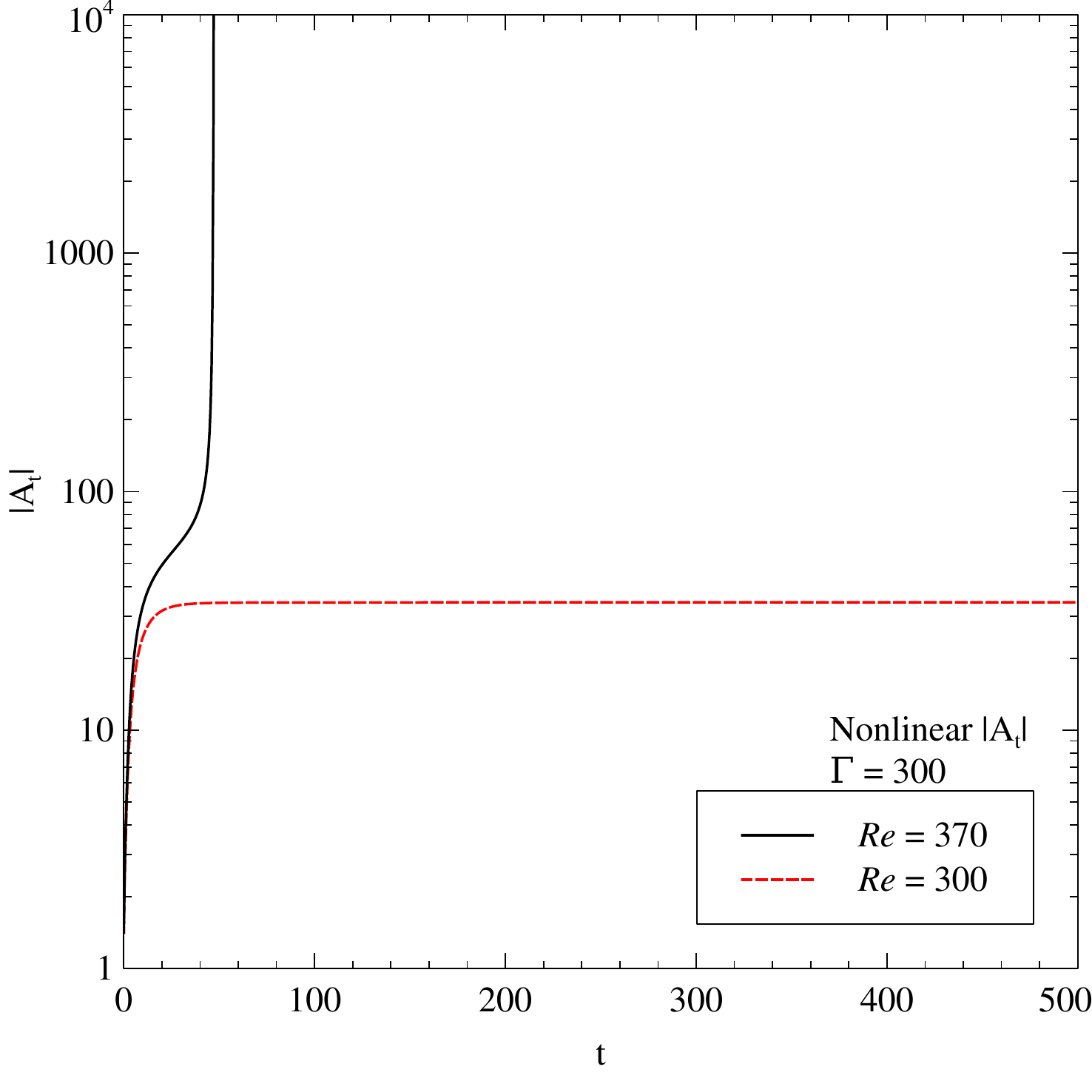}
\caption{Variation of $|A_t|$ as a function of $t$ with $k_y = k_z=1$ for nonlinear analysis in 
plane Couette flow for $\Gamma=300$, and $Re=300$ and $370$ at an initial condition RE($A_t$) = IM($A_t$) = 1.}
\label{fig:A_t_vs_t_plane_cou_nonlin_R_300_370_gamma_3_10_2}
\end{figure}

The situation however changes in the presence of force. Fig. \ref{fig:A_t_vs_t_plane_cou_nonlin_R_300_370_force_no_force} shows that for a small initial amplitude of perturbation, only larger $\Gamma$ makes
$|A_t|$ diverging leading to turbulence. In fact, in the absence of force, $|A_t|$ decays with time
very fast for the range of $Re$ which however could lead to turbulence at higher initial amplitude of
perturbation with $Re>324$ shown by \cite{Duguet_2010} in their simulation in the absence of force. 
In fact \cite{Duguet_2010} argued the initial amplitude of perturbation to be sufficiently large to 
trigger transition to turbulence at $Re$ larger than critical value. However, we can put constraint
on the magnitude of $\Gamma$, based on the simulation of \cite{Duguet_2010}. If $|A_t|$ need not
diverge at $Re<324$, from Fig. \ref{fig:A_t_vs_t_plane_cou_nonlin_R_300_370_force_no_force} we can
argue that $\Gamma$ has to be smaller than $10^4$. Perhaps the upper bound of $\Gamma$ may be such 
that only $Re>324$ will lead to diverging $|A_t|$. Keeping this idea in mind, we show in 
Fig. \ref{fig:A_t_vs_t_plane_cou_nonlin_R_300_370_gamma_3_10_2} that for $\Gamma=300$, while
$|A_t|$ diverges in plane Couette flow hence presumably leading to turbulence for $Re=370$, it
saturates without leading to nonlinear regime for $Re=300$. Note that the saturated $|A_t|$ 
is around 30, whereas critical $|A_t|$ for nonlinearity to arise is 115.04 for $\Gamma=300$ and $Re=300$.
Hence, if the numerical simulation by \cite{Duguet_2010} is our guide, then $\Gamma$ for plane
Couette flow should be around 300.

Nevertheless, the numerical simulations did not include extra force explicitly. Hence, it need not
necessarily mimic exactly what happens in nature. Hence, the above mentioned upper bounds of 
initial amplitude of perturbation and force should be considered with caution and just as 
indicative. While by the virtue of direct numerical simulations, they could consider all 
the modes playing role to reveal turbulence, we have considered extra force in the premise
of least stable mode evolution. Hence, both the frameworks appear to be equivalent. Indeed, for the present purpose, we consider magnitude of extra force as a parameter.
Hence, an independent simulation and also laboratory experimental results help us to constrain
the parameter of the model. 

Above results argue that while $\Gamma$ may have upper bound as expected, large $Re$ requires small $\Gamma$ 
to trigger instability and turbulence. As accretion disc $Re$ is very large, a small $\Gamma$ would
suffice therein. 

\section{Discussion}
\label{diss}

Here we compare our results, i.e. the behaviour of the solution of modified Landau equation with force, with the
conventional perturbation evolution through the Landau equation without force.
The nonlinear evolution of amplitude of perturbations in the absence of extra force (i.e. the usual
Landau equation) is given by equation (\ref{eq:original_landau_eq}) or (\ref{eq:landau_eq}) and the solution is given by 
equation 
(\ref{eq:sol_landau_eq}). Depending on the sign (positive/negative)
of $k_1$ and $k_2$, there are four different possible evolutions of $|A|$ (\citealt{2004hyst.book.....D, 
2002ApMRv..55B..57S}). In the 
present 
context of shear flows, $k_1$ (i.e. $\sigma_i$) is negative, but $k_2$ is positive. Therefore, there will
be a threshold for initial amplitude $A_i$, as shown in equation (\ref{eq:threshold_A}), determining the growth of 
perturbation.
If the initial amplitude $A_0<A_i$, then 
\begin{eqnarray}
|A|^2\sim\frac{A_i^2A_0^2e^{k_1 t}}{A_i^2-A_0^2}
\end{eqnarray} 
at a large $t$. Therefore, $|A|^2\rightarrow 0$ for $A_0<A_i$ at $t\rightarrow\infty$. However, if $A_0>A_i$,
then $|A|^2\rightarrow\infty$ at $t\rightarrow \ln(1-A_i^2/A_0^2)/k_1$.

If both $k_1$ and $k_2$ would be positive, $|A|^2$ blows up after a finite time, given by equation (\ref{eq:t}).
Hence, there will be a fast transition to turbulence. On the other hand, if $k_1>0$ but $k_2<0$,
then $|A|^2\rightarrow k_1/|k_2|$ at $t\rightarrow\infty$. In this case, $|A|^2$ at a large $t$ 
does not depend on $A_0$. Obviously for $k_1$ and $k_2$ both negative, $|A|^2$
decays fast.

However, we have shown in \S\ref{sec:Results} that the saturation in $|A_t|$ is at $|\mathcal{N}|/|\sigma_i|$ in the 
linear regime. We have 
also shown that the assumption of linear analysis at the  saturation of $|A_t|$ may no longer be valid depending on $Re$ 
and $\Gamma$ 
and, hence, the system may already be in the nonlinear regime.  The evolution of $|A_t|$ at the linear regime in our 
case, i.e. with extra force, is similar to 
that of 
$|A|$ from equation (\ref{eq:landau_eq}), i.e. without force, for $k_1>0$ and $k_2<0$. From Figs.~\ref{fig:N_vs_t_R_500_1000_2000} and 
\ref{fig:R_vs_sigma_i_kep}, it is obvious that $|\mathcal{N}|$ increases and $|\sigma_i|$ decreases with the increment of 
$Re$. Therefore, 
at large $Re$ ($\gtrsim 10^{14}$, which is true for accretion discs, see, e.g.,
\citealt{2013PhLB..721..151M}), the saturation of $|A_t|$ is also large and, hence, at 
smaller $\Gamma$ also nonlinearity is inevitable fate of the fluid at the local regime of the accretion disc.  

In the Keplerian and plane Couette flows, $k_1$, i.e. $\sigma_i$, is negative, but $k_2$, i.e. $p_r$, 
could be positive. In the presence of extra force, Landau equation modifies in such a way that the solution
in the linear regime itself mimics the Landau equation without force (i.e. equation \ref{eq:landau_eq}), however,
with $k_1>0$ and $k_2<0$. Further in the nonlinear regime, the amplitude $A_t$ (i.e. with extra force
included) diverges beyond a certain time, depending on $Re$ and $\Gamma$. 
In nonlinear regime, the Landau equation in the presence of extra
force but negative $k_1$ ($\sigma_i$) is, therefore, mimicking the Landau equation
without force but with positive $k_1$ and $k_2$. Essentially, the extra
force effectively changes
the sign of $k_1$ (i.e. $\sigma_i$) for the Landau equation
without force. Speaking in another way, the very presence of extra force destabilizes the 
otherwise stable system.

It is important to note that rotational (Coriolis) effect stabilizes the flow
(see, e.g., \citealt{2005ApJ...629..383M}). Hence, for each $q$, there is an optimum set of
$k_y$ and $k_z$, giving rise to the best least stable mode and growth, which (underlying $\sigma_i$)
decreases with decreasing $q$ below $2$. However, 
at present, we do not concentrate on this feature and in place of optimum set(s) 
of $k_y$ and $k_z$, flows are considered for fixed sets of $k_y$ and $k_z$. Therefore, 
stabilizing effect with respect to rotation does not appear here.

\section{Conclusion}
\label{sec:Conclusion}

Origin of hydrodynamical instability and plausible turbulence in Rayleigh stable flows, e.g. the Keplerian 
accretion disc flow, plane Couette flow, is a long standing problem. While such flows are evident to be
turbulent, they are linearly stable for any Reynolds number. Over the years, several attempts are made
to resolve the problem, with a very limited success, and often the resolution arises with a caveat. 
The major success however in this line lies with MRI, hence in the presence of magnetic field. However,
several astrophysical and laboratory systems are cold, neutral in charge and unmagnetised. Hence, 
any instability therein must be hydrodynamical not magnetohydrodynamical.

We show that in the presence of extra force, governed due to, e.g., thermal fluctuation, grain-fluid interactions,
the amplitude of perturbation may in fact grow with time. Essentially we have established the Landau equation 
for nonlinear perturbation in the presence of Coriolis and external forces. Under suitable combination of 
$Re$ and the external force, perturbation amplitude could be very large. In the linear regime, eventually the amplitude 
saturates beyond a certain time, but the saturated value could be very large, already leading
the system to nonlinear regime, depending on $Re$ (which
is basically controlling the
value of imaginary part of the eigenvalue of perturbation mode) and the force magnitude. In the nonlinear regime,
however, the perturbation amplitude diverges depending on $Re$ and force magnitude. This feature is 
shown to exist in all the apparently Rayleigh stable flows including accretion discs. Thus, the presence 
of force plays an important role to develop nonlinearity and turbulence. As argued here and
in previous literature (e.g. \citealt{2016ApJ...830...86N}), the presence of such force is obvious
and hence hydrodynamical instability and turbulence is not to be a big surprise therein.
Now it is important to confirm the present findings based on direct numerical simulations, which
we plan to undertake in future.

\section*{acknowledgment}

We thank Sujit Kumar Nath of RRI for discussion at the various phases of the work.
Grateful thanks are also due to Jayanta K. Bhattacharjee of IACS, Sandip K.
Chakrabarti of ICSP, Subroto Mukerjee of IISc and Sriram
Ramaswamy of IISc for discussion and suggestions. We are also thankful to Dwight Barkley of the University of 
Warwick and Laurette S. Tuckerman of the Centre national de la recherche scientifique for insightful suggestions and 
fruitful 
discussion, and Srishty Aggarwal of IISc
for giving an independent reading the manuscript and comments for improving the presentation. Finally, last but not least, we thank the referee for an insightful
report and suggestions to improve the presentation of the work.
This work is partly supported by a fund of Department of Science and Technology (DST-SERB) 
with research Grant No. DSTO/PPH/BMP/1946 (EMR/2017/001226).



\section*{Data availability}

No new data were generated or analysed in support of this research.

\bibliographystyle{mnras}
\bibliography{mod_landau_eq_new} 



\onecolumn

\appendix

\section{Modification of background flow in the presence of force}
\label{Modification of background flow in presence of force}
\begin{figure}
 \includegraphics[width=10cm]{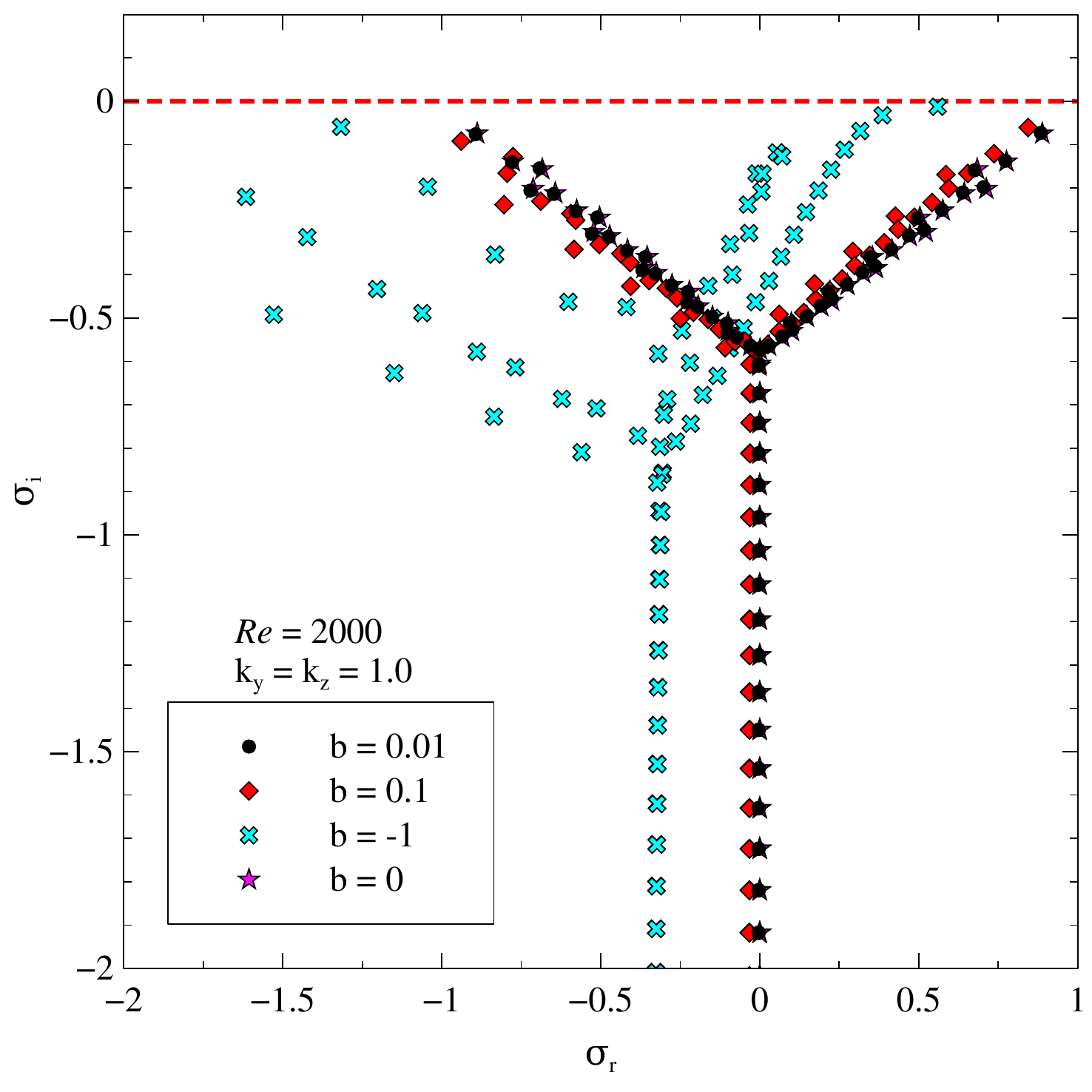}
 \caption{Eigenspectra for background flow $ax+bx^2$ for $Re = 2000$ and $k_y = k_z = 1$ in the 
presence of Coriolis force. Here 
$a = -1$ with different $b$. Note the cases for $b=0$ and $b=0.01$ almost overlap each other.}
 \label{fig:eval_cou_poi_rot_diff_b_3d}
\end{figure}

Due to the presence of the extra force, the background flow may be modified 
from its plane Couette flow nature. Let us understand it from a simplistic consideration.
Considering the background flow 
\begin{eqnarray}
\label{vbk}
\bfv= (0, V_Y(X), 0),
\end{eqnarray} 
the Navier-Stokes equation in the presence of force is
\begin{eqnarray}
 \frac{\partial \bfv}{\partial t} + (\bfv \cdot \nabla)\bfv = -\frac{\nabla P}{\rho} +\nu\nabla^2 \bfv + \bm{F},
 \label{eq:N_S_equ}
\end{eqnarray}
where $P$, $\rho$, $\nu$ and $\textbf{F}$ are the pressure, density, kinematic viscosity and extra force, chosen constant 
for the present purpose, respectively. 
The three components of equation (\ref{eq:N_S_equ}) are
\begin{eqnarray}
 0 = -\frac{1}{\rho}\frac{\partial P}{\partial X} + F_X,
 \label{eq:x_comp}
\end{eqnarray}
\begin{eqnarray}
\begin{split}
 0 = -\frac{1}{\rho}\frac{\partial P}{\partial Y} + \nu \nabla^2 V_Y + F_Y,
 \label{eq:y_compp}
 \end{split}
\end{eqnarray}
\begin{eqnarray}
 0 = -\frac{1}{\rho}\frac{\partial P}{\partial Z} + F_Z.
 \label{eq:x_comp}
\end{eqnarray}
Equation (\ref{eq:y_compp}) can be further simplified to 
\begin{eqnarray}
\nabla^2 V_Y = \frac{1}{\nu}(-F_Y+\frac{1}{\rho}\frac{\partial P}{\partial Y}) = \frac{\partial^2 
V_Y}{\partial X^2}.
\end{eqnarray}
Hence, for constant $\partial P/\partial Y$ and $F_Y$,
\begin{eqnarray}
\begin{split}
V_Y = -\left(\frac{F_Y}{\nu}-\frac{1}{\nu\rho}\frac{\partial P}{\partial Y}\right)\frac{X^2}{2} + 
C_1X+C_2& \\
 = -K\frac{X^2}{2} + C_1X + C_2, 
\label{bgk}
\end{split}
\end{eqnarray}
where 
\begin{eqnarray}
K = \left(\frac{F_Y}{\nu}-\frac{1}{\nu\rho}\frac{\partial 
P}{\partial Y}\right).
\end{eqnarray}
The corresponding boundary conditions
\begin{eqnarray}
 V_Y = \mp U_0\ {\rm at}\  X=\pm L
\end{eqnarray}
lead $V_Y$ in equation (\ref{bgk}) to 
\begin{eqnarray}
 V_Y = \frac{K}{2}(L^2-X^2)-\frac{U_0X}{L}, 
 \label{eq:PPF_reduction}
\end{eqnarray}
where $F_Y=\partial P/\partial Y=0$ brings the background back to plane Couette/shear flow.
In fact, for ideal plane Couette flow, there is no pressure gradient along any direction. 
Therefore, $K$ becomes $F_Y/\nu$ only and $F_X=F_Z=0$, assuring the choice of equation (\ref{vbk}). 
In accretion discs, however, $Re$ is very large (and $\nu$ is very small). Hence for a given
$K$, a very small $F_Y$ suffices. In fact, it has been shown in \S \ref{sec:lin-ana} that with
the increase of $Re$, $\Gamma$ has to be increasingly small in order to maintain linear
approach intact. Therefore, $F_Y$ can be smaller than smallness of $\nu$ and therefore the 
effect of nonlinear term in equation (\ref{eq:PPF_reduction}) is very small. 
The flow, therefore, effectively becomes plane Couette flow (or the Keplerian 
flow in the presence of rotation/Coriolis effect) only. Fig.~\ref{fig:eval_cou_poi_rot_diff_b_3d} 
shows the eigenspectra for the background flow of form $ax+bx^2$, where $x=X/L$, the dimensionless
length. See Appendix \ref{sec:derivation of OS and Sqiuire eqs} for all details of the units. 
This background flow mimics that given by equation (\ref{eq:PPF_reduction}). 
However, we observe that the small 
value of $b/a$ does not affect the eigenspectra much and it almost remains the same as
of the Keplerian flow. As small $b/a$ corresponds to small $F_Y$, we can assume background
flow of linear shear in our model calculations throughout, particularly for high $Re$ 
flows, e.g. Keplerian flow which is the central essence of the work, 
where indeed force to be very small (see \S \ref{sec:lin-ana}).
However, following \cite{Farrell_1993}, we also can assume that extra force arises only due to 
perturbation. Hence, the background flow remains intact, the same as linear shear flow.
This situation has been explored in Appendix \ref{sec:derivation of OS and Sqiuire eqs}.


\section{Derivation of Orr-Sommerfeld and Squire equations in the presence of Coriolis and external forces} 
\label{sec:derivation of OS and Sqiuire eqs}

Let us consider small shearing box centered at the radius $r_0$ with angular velocity $\Omega_0=U_0/qL$ 
with size in $x-$direction 
2$L=r-r_0$, and we are 
going to observe the motion of 
the fluid with respect to that box.
The model is described in \citealt{2005ApJ...629..383M}. The unperturbed velocity for linear shear is 
\begin{eqnarray}
 \textbf{V} = \left(0,-\frac{U_0 X}{L},0\right),
 \label{eq:velo_prof}
\end{eqnarray}
where $X$ is dimensionful $x$-coordinate.
Again the angular velocity vector $\bm{\omega} = (0,0,\Omega_0)$ when $\Omega=\Omega_0\left(r_0/r\right)^q$.
Now to study the dynamics of a viscous and incompressible rotating fluid, let us consider Navier-Stokes equation in the 
presence of Coriolis and centrifugal forces, i.e.
\begin{eqnarray}
\begin{split}
 \frac{\partial \textbf{V}}{\partial t'} + (\textbf{V}\cdot \bm{\nabla}')\textbf{V} = -\frac{1}{\rho}\bm{\nabla}'P - 
\bm{\omega}\times\bm{\omega}\times\textbf{D}-2\bm{\omega}\times\textbf{V}  + \nu \nabla'^2\textbf{V}
 \label{eq:N-S_eq}
 \end{split}
\end{eqnarray}
and continuity equation, i.e.
\begin{eqnarray}
 \bm{\nabla}'\cdot \textbf{V} = 0,
 \label{eq:continuity_eqn}
\end{eqnarray}
where $P$ and $\rho$ are the pressure and density of the fluid respectively.

Here $\textbf{D} = (X,Y,Z)$, $\bm{\nabla}' = \left(\frac{\partial}{\partial X},\frac{\partial}{\partial 
Y},\frac{\partial}{\partial 
Z}\right)$.
To express the above equations in dimensionless variables, we define
$$X=xL,\ Y = yL,\ Z= zL,\ \textbf{V}=U_0 \textbf{U},\ t' = \frac{tL}{U_0},\ \textbf{U} = (0,-x,0).$$

Now equation (\ref{eq:N-S_eq})  becomes
\begin{eqnarray}
\begin{split}
 \frac{\partial \textbf{U}}{\partial t} + (\textbf{U}\cdot 
\bm{\nabla})\textbf{U}+\frac{1}{q^2}(\hat{k}\times\hat{k}\times\textbf{d}) 
+\frac{2\hat{k}\times\textbf{U}}{q} +\bm{\nabla}\bar{p} = \frac{1}{Re} \nabla^2\textbf{U}
 \label{eq:N-S_eqn_dim_less}
 \end{split}
\end{eqnarray}
and equation (\ref{eq:continuity_eqn}) becomes
\begin{eqnarray}
 \bm{\nabla}\cdot \textbf{U} = 0.
 \label{eq:continuity_eqn_dim_less}
\end{eqnarray}

Here $\bar{p} = \frac{P}{U_0^2 \rho},\ \textbf{d} = (x,y,z),\ \bm{\nabla} = \left(\frac{\partial}{\partial 
x},\frac{\partial}{\partial 
y},\frac{\partial}{\partial z}\right)$ and $Re = \frac{U_0 L}{\nu}$.

 Now we perturb the system and as a result $\textbf{U}(x)\rightarrow \textbf{U}(x)+\textbf{u}'(x,y,z,t)$ and 
$\bar{p}\rightarrow \bar{p} +p'(x,y,z,t)$, where $\textbf{u}' = (u,v,w)$. Due to the perturbation an 
extra stochastic force, $\bm{F}(x,y,z,t)$, will arise in the system, as argued by \cite{Farrell_1993}.
Also following Appendix \ref{Modification of background flow in presence of force}, we can neglect 
any effect of force before perturbation.

Hence the evolution equation of perturbation from perturbed 
Navier-Stokes equation is given by
\begin{eqnarray}
\begin{split}
 \frac{\partial \textbf{u}'}{\partial t} + (\textbf{U}\cdot 
\bm{\nabla})\textbf{u}'+(\textbf{u}'\cdot 
\bm{\nabla})\textbf{U} +\frac{2\hat{k}\times\textbf{u}'}{q} +\bm{\nabla}p' = \frac{1}{Re} 
\nabla^2\textbf{u}'\ -(\textbf{u}'\cdot \bm{\nabla})\textbf{u}'
+\bm{F}(x,y,z,t)
 \label{eq:N-S_eqn_linear}
 \end{split}
\end{eqnarray}
and continuity equation becomes 
\begin{eqnarray}
 \bm{\nabla}\cdot \textbf{u}' = 0,
 \label{eq:continuity_eqn_linear}
\end{eqnarray}
when $\bm{F}$ is the final stochastic force.
For our convenience we use the following notation:
$$\textbf{U} = (0,U_y,0); \ U_y = U(x) = -x.$$
Componentwise equations (\ref{eq:N-S_eqn_linear}) and (\ref{eq:continuity_eqn_linear}) become
\begin{eqnarray}
\left(\frac{\partial}{\partial t}+U_y \frac{\partial}{\partial y}\right)u - \frac{2v}{q}+\frac{\partial p'}{\partial x} 
=& 
\frac{1}{Re} 
\nabla^2 u\ -(\textbf{u}'\cdot \bm{\nabla})u +F_x
\label{eq:x_comp}\\
\left(\frac{\partial}{\partial t}+U_y \frac{\partial}{\partial y}\right)v + u\frac{\partial U_y}{\partial x} + 
\frac{2u}{q}+\frac{\partial 
p'}{\partial y} =& \frac{1}{Re} \nabla^2 v \ -(\textbf{u}'\cdot \bm{\nabla})v +F_y
\label{eq:y_comp}\\
\left(\frac{\partial}{\partial t}+U_y \frac{\partial}{\partial y}\right)w +\frac{\partial p'}{\partial z} =& \frac{1}{Re} 
\nabla^2 w \ 
-(\textbf{u}'\cdot \bm{\nabla})w +F_z
\label{eq:z_comp}\\
\frac{\partial u}{\partial x}+\frac{\partial v}{\partial y}+\frac{\partial w}{\partial z} = 0,
\label{eq:continuity_eqn_derivative}
\end{eqnarray}
where the $x$-component of vorticity is $\zeta = \left(\frac{\partial w}{\partial y} - \frac{\partial v}{\partial 
z}\right)$.

We further take divergence on both the sides of equation (\ref{eq:N-S_eqn_linear}) and exploit equation 
(\ref{eq:continuity_eqn_linear}) to 
have 
\begin{eqnarray}
\bm{\nabla}\cdot\left\{(\textbf{U}\cdot 
\bm{\nabla})\textbf{u}'\right\}+\bm{\nabla}\cdot\left\{(\textbf{u}'\cdot 
\bm{\nabla})\textbf{U}\right\} +\bm{\nabla}\cdot\frac{2\hat{k}\times\textbf{u}'}{q} +\nabla^2 p' =
-\bm{\nabla} \cdot [(\textbf{u}'\cdot \bm{\nabla})\textbf{u}'] +\bm{\nabla}\cdot\bm{F},
\end{eqnarray}
where
\begin{eqnarray*}
        \bm{\nabla}\cdot\left\{(\textbf{U}\cdot \bm{\nabla})\textbf{u}'\right\} &=& \bm{\nabla}\cdot \left(U_y 
\frac{\partial 
\textbf{u}'}{\partial y}\right)\\
         &=&\bm{\nabla} U_y \cdot \frac{\partial \textbf{u}'}{\partial y} + U_y \frac{\partial}{\partial 
y}(\bm{\nabla}\cdot\textbf{u}')\\
        &=& \frac{\partial U(x)}{\partial x} \frac{\partial u}{\partial y},
       \end{eqnarray*}
\begin{eqnarray*}
        \bm{\nabla}\cdot\left\{(\textbf{u}'\cdot \bm{\nabla})\textbf{U}\right\}  = \frac{\partial u}{\partial y} 
\frac{\partial 
U(x)}{\partial x},
       \end{eqnarray*}
\begin{eqnarray*}
        \bm{\nabla} \cdot \left\{\hat{k}\times\textbf{u}'\right\} = -\frac{\partial v}{\partial x}+\frac{\partial 
u}{\partial y}.
       \end{eqnarray*}
Hence, 
\begin{eqnarray}
 \nabla^2 p' = -\bm{\nabla} \cdot [(\textbf{u}'\cdot 
\bm{\nabla})\textbf{u}'] - 2 \frac{\partial 
U(x)}{\partial x} \frac{\partial u}{\partial y} + \frac{2}{q}\left(\frac{\partial v}{\partial 
x}-\frac{\partial u}{\partial y}\right) + \bm{\nabla}\cdot \bm{F}.
  \label{eq:pressure_eqn}
\end{eqnarray}
If we take gradient and then divergence in equation (\ref{eq:x_comp}) and also use equation 
(\ref{eq:pressure_eqn}), we obtain 
\begin{eqnarray}
\begin{split}
 \left(\frac{\partial}{\partial t}+U\frac{\partial}{\partial y}\right)\nabla^2u -\frac{\partial^2 U}{\partial x^2} 
\frac{\partial 
u}{\partial y}+ \frac{2}{q}\frac{\partial \zeta}{\partial z} = \frac{1}{Re}\nabla^4u\ +NL^u  +
F_1,
\label{eq:orr_sommerfeld_eq_before_ensm_avg}
\end{split}
\end{eqnarray}
and if we do partial derivatives with respect to $y$  of equation (\ref{eq:z_comp}) and with respect to $z$ of the 
equation (\ref{eq:y_comp}), and subtract one from the other, we end up with
\begin{eqnarray}
 \left(\frac{\partial}{\partial t}+U\frac{\partial}{\partial y}\right)\zeta -\left(\frac{\partial U}{\partial x} + 
\frac{2}{q}\right) 
\frac{\partial u}{\partial z} = \frac{1}{Re}\nabla^2 \zeta +NL^{\zeta} + 
F_2,
\label{eq:squire_eq_before_ensm_avg}
\end{eqnarray}
where $F_1$, $F_2$, $NL^u$ and $NL^{\zeta}$ are given by

\begin{eqnarray}
 F_1 = \left(\nabla^2 F_x - \bm{\nabla}\cdot \frac{\partial \bm{F}}{\partial x}\right),
 \label{eq:F_1}\\
 F_2 = \left(\frac{\partial F_z}{\partial y} - \frac{\partial F_y}{\partial z}\right),
 \label{eq:F_2}\\
NL^u = -\nabla^2\cdot [(\textbf{u}'\cdot \bm{\nabla})u]+\bm{\nabla} \cdot \frac{\partial}{\partial x} 
[(\textbf{u}'\cdot \bm{\nabla})\textbf{u}'],
\label{eq:NLu}
\\
NL^{\zeta} = - \frac{\partial}{\partial y}[(\textbf{u}'\cdot \bm{\nabla})w]+ \frac{\partial}{\partial 
z}[(\textbf{u}'\cdot \bm{\nabla})v].
\label{eq:NLzeta}
\end{eqnarray}

If $F_1$ and $F_2$ happen to be Gaussian in nature, their ensemble average in the present context of biased stochastic 
system turns out to be non-zero constants, respectively $\Gamma_1$ and $\Gamma_2$, appeared in equations 
(\ref{eq:orr_sommerfeld_eq}) and 
(\ref{eq:squire_eq}). 


\section{Nonlinear terms for threedimensional perturbation}
\label{The nonlinear terms for three dimensional perturbations}
Here we show how to calculate the coefficient of $|A_t|^2A_t$ in equation (\ref{eq:mod_landau_eq}) (i.e. $p$). The 
nonlinear terms 
corresponding to equation (\ref{eq:1st_nonlin_orr_sommerfeld_eq}) and (\ref{eq:1st_nonlin_squire_eq}) are
\begin{eqnarray}
\begin{split}
 NL^u_1 = i\frac{\partial}{\partial x}\Bigl[\bar{u}^*_1\frac{\partial}{\partial x}(k_y \bar{v}_2+k_z \bar{w}_2) + 2ik_y 
\bar{v}^*_1 (k_y 
\bar{v}_2+k_z \bar{w}_2) + 2ik_z \bar{w}^*_1 (k_y \bar{v}_2+k_z \bar{w}_2)\Bigr] \\+ k^2\Bigr[\bar{u}^*_1\frac{\partial 
\bar{u}_2}{\partial x}  + 2i\bar{u}_2 (k_y \bar{v}^*_1+k_z \bar{w}^*_1)\Bigl] \\ + i\frac{\partial}{\partial 
x}\Bigl[\bar{u}_2 
\frac{\partial}{\partial x}(k_y \bar{v}^*_1+k_z \bar{w}^*_1)  - ik_y\bar{v}_2(k_y \bar{v}^*_1+k_z \bar{w}^*_1) -ik_z 
\bar{w}_2 (k_y 
\bar{v}^*_1+k_z \bar{w}^*_1)\Bigr] \\ +k^2\Bigr[\bar{u}_2\frac{\partial \bar{u}^*_1}{\partial x} -i\bar{u}^*_1 (k_y 
\bar{v}_2+k_z 
\bar{w}_2)\Bigl]
\\
= i\frac{\partial}{\partial x}\Bigl[\bar{u}^*_1\frac{\partial}{\partial x}(k_y \bar{v}_2+k_z \bar{w}_2) +\bar{u}_2 
\frac{\partial}{\partial x}(k_y \bar{v}^*_1+k_z \bar{w}^*_1) + i(k_y \bar{v}^*_1+k_z \bar{w}^*_1)(k_y \bar{v}_2+k_z 
\bar{w}_2)\Bigr] \\+ 
k^2\Bigr[\frac{\partial}{\partial x}(\bar{u}^*_1 \bar{u}_2) + 2i\bar{u}_2(k_y \bar{v}^*_1+k_z \bar{w}^*_1) 
-i\bar{u}^*_1(k_y 
\bar{v}_2+k_z \bar{w}_2)\Bigr]
\end{split}
\end{eqnarray}
and 
\begin{eqnarray}
\begin{split}
NL^{\zeta}_1 = -ik_y\left(\bar{u}_2\frac{\partial \bar{w}^*_1}{\partial x} + \bar{u}^*_1 \frac{\partial 
\bar{w}_2}{\partial x}\right) + 
k_y^2(2\bar{v}^*_1 \bar{w}_2 - \bar{v}_2 \bar{w}^*_1) + k_y k_z \bar{w}^*_1 \bar{w}_2 \\+ ik_z 
\left(\bar{u}_2\frac{\partial 
\bar{v}^*_1}{\partial x} + \bar{u}^*_1 \frac{\partial \bar{v}_2}{\partial x}\right) - k_z^2 (2\bar{w}^*_1 \bar{v}_2 - 
\bar{w}_2 
\bar{v}^*_1) - k_y k_z \bar{v}^*_1 \bar{v}_2.
\end{split}
\end{eqnarray}

From equations (\ref{eq:Q_1}), (\ref{eq:Q_1_trial_sol}) and (\ref{eq:Q_x}) we have
\begin{eqnarray}
\bar{u}_1(x,t) = A_t \phi^u, \ \  \bar{\zeta}_1(x,t) = A_t \phi^{\zeta},
\label{eq:u_1_zeta_1}
\end{eqnarray}
where we have neglected $\frac{\partial}{\partial x}\Bigl(\frac{1}{\mathcal{D}_t+i\mathcal{L}}\Bigr)\Gamma$ 
over $\frac{\partial}{\partial t}\Bigl(\frac{1}{\mathcal{D}_t+i\mathcal{L}}\Bigr)\Gamma$, 
where the former is expected to be much smaller than the latter numerically.
In otherwords, we compute nonlinear effects based on the homogeneous part 
of ($\bar{u}_1$, $\bar{v}_1$, $\bar{w}_1$) and ($\bar{u}_2$, $\bar{v}_2$, $\bar{w}_2$), and their complex 
conjugates, to the first approximation
in equations (\ref{eq:1st_nonlin_orr_sommerfeld_eq}) and (\ref{eq:1st_nonlin_squire_eq}).

In general, once $u$ and $\zeta$ are known from equations (\ref{eq:orr_sommerfeld_eq}) and (\ref{eq:squire_eq}) 
(irrespective of inclusion 
of 
noise/force and nonlinear terms), the other two components of the perturbed velocity field can be obtained from equations
\begin{eqnarray}
 -\frac{\partial u}{\partial x} &=& \frac{\partial v}{\partial y} + \frac{\partial w}{\partial z},\\
 \zeta &=& -\frac{\partial v}{\partial z} + \frac{\partial w}{\partial y}
 \label{eq:continuity_vorticity_eq}
\end{eqnarray}
and the governing equations are 
\begin{eqnarray}
\Bigl(\frac{\partial^2}{\partial z^2} + \frac{\partial^2}{\partial y^2}\Bigr)w = - \frac{\partial^2 u}{\partial z \partial 
x} 
+ \frac{\partial \zeta}{\partial y},
\label{eq:v_w_2eqs}
\end{eqnarray}
\begin{eqnarray}
\Bigl(\frac{\partial^2}{\partial z^2} + \frac{\partial^2}{\partial y^2}\Bigr)v = - \frac{\partial^2 u}{\partial y \partial 
x} 
- \frac{\partial \zeta}{\partial z}.
\label{eq:v_w_eqs}
\end{eqnarray}
Note, $\bar{v}_1$ and $\bar{w}_1$ have the same time dependence as $\bar{u}_1$ or $\bar{\zeta}_1$ has, as the above two 
equations do not 
contain 
time derivative explicitly. They, therefore, can be written as
\begin{eqnarray}
 \bar{v}_1 = A_t \phi^v,\ \ \bar{w}_1 = A_t \phi^w,
 \label{eq:v_1_w_1}
\end{eqnarray}
where 
\begin{eqnarray}
 \phi^v &=& \frac{ik_y}{k^2} \frac{d \phi^u}{d x} + \frac{ik_z}{k^2}\phi^{\zeta}
\end{eqnarray}
and
\begin{eqnarray}
 \phi^w &=& \frac{ik_z}{k^2} \frac{d \phi^u}{d x} - \frac{ik_y}{k^2}\phi^{\zeta},
\end{eqnarray}
on substituting $u,\zeta=A_t\phi^{u,\zeta}e^{i\textbf{k}\cdot\textbf{r}}$ with
$\textbf{k}=(0,k_y,k_z)$ and $\textbf{r}=(0,y,z)$ in equations 
(\ref{eq:v_w_2eqs}) and (\ref{eq:v_w_eqs}).

The calculations for $u_2$, $v_2$ and $w_2$ are shown in \citealt{2011MNRAS.414..691R}.

\section{Determination of $\Gamma'$}
\label{sec:The calculation of Gamma}
From equation (\ref{eq:exp_for_Gamma'}) we have
\begin{eqnarray}
\begin{split}
\Gamma' =&-\Gamma + i\sigma_r\Bigl(\frac{1}{\mathcal{D}_t+i\mathcal{L}}\Bigr)\Gamma.
\label{eq:exp_for_Gamma'_only}
\end{split}
\end{eqnarray}
The second term is
\begin{eqnarray}
 \frac{1}{\mathcal{D}_t+i\mathcal{L}}\Gamma = \frac{\mathcal{D}_t-i\mathcal{L}}{\mathcal{D}_t^2+\mathcal{L}^2}\Gamma = 
\frac{\mathcal{D}_t}{\mathcal{D}_t^2+\mathcal{L}^2}\Gamma - \frac{i\mathcal{L}}{\mathcal{D}_t^2+\mathcal{L}^2}\Gamma.
\end{eqnarray}
Now, we have
\begin{eqnarray}
\frac{\mathcal{D}_t}{\mathcal{D}_t^2+\mathcal{L}^2}\Gamma = \mathcal{D}_t^{-2} \mathcal{D}_t 
\Bigl(1+\frac{\mathcal{L}^2}{\mathcal{D}_t^2}\Bigr)^{-1}\Gamma.
\label{dtterm}
\end{eqnarray}
If $||\mathcal{D}_t^2||>||\mathcal{L}^2||$, then the R.H.S. of equation (\ref{dtterm}) can be written as
\begin{eqnarray}
\begin{split}
\mathcal{D}_t^{-1} \Bigl(1-\frac{\mathcal{L}^2}{\mathcal{D}_t^2}+\frac{\mathcal{L}^4}{\mathcal{D}_t^4} - 
...\Bigr)\Gamma\\=\Bigl(\frac{1}{\mathcal{D}_t}-\frac{\mathcal{L}^2}{\mathcal{D}_t^3}+\frac{\mathcal{L}^4}{\mathcal{D}_t^5
} - 
...\Bigr)\Gamma\\ = \Bigl(t-\mathcal{L}^2t^3+\mathcal{L}^4t^5 - ...\Bigr)\Gamma\\ = 
t\Bigl(1-\mathcal{L}^2t^2+\mathcal{L}^4t^4 - 
...\Bigr)\Gamma\\ = t\Bigl(1+\mathcal{L}^2t^2\Bigr)^{-1}\Gamma.
\end{split}
\label{dtinvop}
\end{eqnarray}
Similarly, we have the other term
\begin{eqnarray}
\begin{split}
\frac{i\mathcal{L}}{\mathcal{D}_t^2+\mathcal{L}^2}\Gamma = 
\frac{i\mathcal{L}}{\mathcal{D}_t^2}\Bigl(1+\frac{\mathcal{L}^2}{\mathcal{D}_t^2}\Bigr)^{-1}\Gamma \\ = 
\frac{i\mathcal{L}}{\mathcal{D}_t^2} \Bigl(1-\frac{\mathcal{L}^2}{\mathcal{D}_t^2}+\frac{\mathcal{L}^4}{\mathcal{D}_t^4} 
- 
...\Bigr)\Gamma\\ = i\mathcal{L}t^2(1-\mathcal{L}^2t^2 + \mathcal{L}^4t^4 - ...)\Gamma \\= 
i\mathcal{L}t^2(1+\mathcal{L}^2t^2)^{-1}\Gamma.
\end{split}
\end{eqnarray}
Hence
\begin{eqnarray}
\Gamma' = -\Gamma +i\sigma_r(t-i\mathcal{L}t^2)(1+\mathcal{L}^2t^2)^{-1}\Gamma.
\end{eqnarray}
When $t$ is large, the above expression becomes
\begin{eqnarray}
\Gamma' = -\Gamma +i\sigma_r\Bigl(\frac{-i}{\mathcal{L}}\Bigr)\Gamma = -\Gamma 
+\sigma_r\Bigl(\frac{1}{\mathcal{L}}\Bigr)\Gamma.
\label{eq:Gamma'_large_t}
\end{eqnarray}

\bsp	
\label{lastpage}
\end{document}